\documentclass[useAMS,usenatbib]{mn2e}
\usepackage{subfigure}
\usepackage{graphicx}

\title[Nuclear activity versus star formation at $z\sim0.8$]{Probing nuclear activity versus star formation at $z\sim0.8$ using 
near-infrared multiobject spectroscopy}
\author[C. Ramos Almeida et al.]
{\parbox{\textwidth}{C. Ramos Almeida$^{1,2}$\thanks{E-mail:cra@iac.es},
J. M. Rodr\'\i guez Espinosa$^{1,2}$,
J. A. Acosta-Pulido$^{1,2}$,
A. Alonso-Herrero$^{3,4}$, 
A. M. P\'erez Garc\' ia$^{1,2}$,
N. Rodr\' iguez-Eugenio$^{1,2}$
}\vspace{0.4cm}\\
\parbox{\textwidth}{$^{1}$Instituto de Astrof\'\i sica de Canarias (IAC), C/V\'\i a L\'{a}ctea, s/n, E-38205, La Laguna, 
Tenerife, Spain\\
$^{2}$Departamento de Astrof\' isica, Universidad de La Laguna, E-38205, La Laguna, Tenerife, Spain\\
$^{3}$Instituto de F\' isica de Cantabria, CSIC-Universidad de Cantabria, E-39005 Santander, Spain\\
$^{4}$Augusto Gonz\'alez Linares Senior Research Fellow}
}
\begin{document}

\date{}

\pagerange{\pageref{firstpage}--\pageref{lastpage}} \pubyear{2011}

\maketitle

\label{firstpage}

\begin{abstract}
We present near-infrared (NIR) spectroscopic observations of 28 X-ray and mid-infrared selected sources at a 
median redshift of $z\sim0.8$ in the Extended Groth Strip (EGS). To date this is the largest 
compilation of NIR spectra of active galactic nuclei (AGN) at this redshift.  
The data were obtained using the multi-object spectroscopic mode of the Long-slit Intermediate Resolution 
Infrared Spectrograph (LIRIS) at the 4.2 m William Herschel Telescope (WHT).
These galaxies are representative of a larger sample studied in a previous work, consisting of over a hundred 
X-ray selected sources with mid-infrared counterparts, which were classified either as {\it AGN-dominated} or 
{\it host galaxy-dominated}, depending on the shape of their spectral energy distributions (SEDs).
Here we present new NIR spectra of 13 and 15 sources of each class respectively. 
We detect the H$\alpha$ line at $\ge 1.5\sigma$ above the continuum for the majority of the galaxies. 
Using attenuation-corrected H$\alpha$ luminosities and observed Spitzer/MIPS 24 \micron~fluxes, and after subtracting 
an AGN component estimated using an AGN empirical correlation and multifrequency SED fits, 
we obtain average star formation rates (SFRs) of 7$\pm$7 and 20$\pm$50 M$_{\sun}~yr^{-1}$ respectively
(median SFRs = 7 and 5 M$_{\sun}~yr^{-1}$).
These values are lower than the SFRs reported in the 
literature for different samples of non-active star-forming galaxies of similar stellar masses and redshifts 
(M$_*\sim 10^{11}M_{\sun}$ and $z\sim$1). In spite of the small size of the sample studied here, as well as the
uncertainty affecting the AGN-corrected SFRs, we speculate 
with the possibility of AGN quenching the star formation in galaxies at $z\sim0.8$. Alternatively, we might be 
seeing a delay between the offset of the star formation and AGN activity, as observed in the local universe.
\end{abstract}


\section{Introduction}

The role of active galactic nuclei (AGN) in the formation and evolution of galaxies is still not well
established. It is not clear whether AGN represent episodic phenomena in the life of galaxies, are random
processes (given that the supermassive black hole is already there, at least in the local universe), 
or are more fundamental. 
Some authors claim that AGN are key in quenching the star formation in their host galaxies 
through the so-called AGN feedback (see e.g. \citealt{Granato04,Ho05,Springel05,Schawinski07,Schawinski09}). 
This ``negative'' AGN feedback has been invoked to explain the well-established correlations between supermassive 
black hole (SMBH) mass and host galaxy properties (e.g. \citealt{Kormendy95,Magorrian98,Ferrarese00,
Gebhardt00,Greene06}). 
%However, the nonetheless well-established correlation
%between the bulge mass and the supermassive black hole (SMBH) mass 
%\citep{Kormendy95,Magorrian98,Ferrarese00,Gebhardt00,Greene06} tells us that very likely galaxy 
%bulges and SMBHs formed at the same time and through similar processes. 
A similar correlation has been observed in starburst galaxies, 
in which bright stellar clusters, also called Central
Massive Objects (CMO), take the role of the SMBH \citep{Ferrarese06,Wehner06}. 
%It has also 
%been shown that the mass dependence of the peak star formation epoch appears to mirror the mass dependence 
%of BH activity, as seen in surveys of both radio- and X-ray-selected active galactic nuclei 
%\citep{Waddington01,Hasinger03}.

In order to understand the importance of the AGN feedback in the evolution and the
star formation histories of galaxies, it is necessary to study how the star formation rate (SFR)
in active galaxies evolve with redshift. 
AGN at cosmological distances have been widely studied in the optical due to the relatively high 
number of multi-object spectrographs in this range. On the other hand, the sparse number of these instruments
in the NIR translates in the lack of spectroscopic studies of distant AGN in the NIR.
%The NIR offers a wide variety of diagnostic tools 
%to characterize relevant phenomena, with the additional advantage of being far less affected
%by extinction than those in the optical and the ultraviolet (UV). 
This range offers the opportunity to study the optical spectra of galaxies at $z\sim1$. 
At this redshift, the H$\alpha$ line is shifted into the J-band. This recombination line is a good 
tracer of the instantaneous SFR, since it is proportional to the ionising UV Lyman continuum radiation
from young and massive stars and it has little dependence on metallicity 
(\citealt{Kennicutt98,Bicker05}). Characterising the SFRs of active galaxies
using either H$\alpha$ or any other indicator (e.g.~the IR emission) is very challenging, 
as the AGN contributes to both the continuum and the emission line spectrum. Thus, estimates of the SFRs of
powerful AGN, as for example quasars, will be contaminated with AGN emission to a certain extent. 
On the other hand, the SFRs of deeply buried and optically-dull AGN obtained
from H$\alpha$ emission may only have a small contribution from the AGN. These optically-dull AGN are defined
as X-ray-selected AGN with no evidence for nuclear accretion activity in optical spectroscopy, showing stellar 
emission-dominated or obscured optical-to-infrared SEDs instead, practically indistinguishable 
from those of spiral/starburst galaxies \citep{Alonso04,Alonso08,Rigby06,Trump09}.

The star formation activity in the hosts of AGN at $z\sim$1 has been studied by 
several authors using mid-infrared (MIR), far-infrared (FIR), and submillimiter data 
\citep{Alonso08,Bundy08,Brusa09,Lutz10,Santini12}.
The latter authors found that the period of moderately luminous AGN activity does not seem to have 
strong influence in the star formation activity of the galaxies, in contradiction with the results 
found at low redshift (e.g.~\citealt{Ho05}). For example, based on FIR data from the 
Herschel Space Observatory, \citet{Santini12} reported evidence of an enhancement on the star formation 
activity in the host galaxies of a sample of X-ray-selected AGN at $0.5<z<2.5$, as compared to a mass-matched 
control sample of non-active galaxies. However, when they only considered star-forming galaxies in the 
control sample (i.e.~they discarded quiescent galaxies), they found roughly the same level of star formation 
as in the AGN hosts. A similar result was found by \citet{Lutz10}, based on submillimeter data of a sample of 
895 X-ray selected AGN at $z\sim$1, for which they measured a SFR$\sim$30 $M_{\sun}~yr^{-1}$. This value, 
which they estimated assuming star formation-dominated submillimeter emission, 
is among the typical SFRs found for samples of non-active star-forming galaxies at $z\sim$1 and 
$M_* \ga 10^{10.5}M_{\sun}$ (e.g. \citealt{Noeske07}). \citet{Alonso08} studied the star formation 
properties of 58 X-ray-selected AGN at $0.5<z<1.4$ by modelling their multifrequency SEDs and did not
found strong evidence for either highly supressed or enhanced star formation when compared to a mass-matched 
sample of galaxies at the same redshift. However, these AGN were selected to have SEDs dominated by 
stellar emission, and thus, they are representative of only 50\% of the X-ray-selected AGN population.

In our previous work (Ramos Almeida et al.~2009; hereafter~\citealt{Ramos09}) we fitted the optical to
MIR SEDs of a sample of 116 X-ray-selected AGN in the Extended Groth Strip (EGS)
with different starburst, AGN, and galaxy templates from \citet{Polletta07}.
Based on this SED fitting, we classified them as AGN-dominated (52\%) and host galaxy-dominated (48\%) objects.
The latter have SEDs typical of starburst, spiral, or elliptical galaxies, indicating the presence of a 
deeply buried or a low-luminosity AGN.
%Thus, at least in the case of these ``optical and NIR dull AGN'', it 
%should be possible to estimate relatively accurate SFRs from their H$\alpha$ luminosities. 
From our SED fits, we derived photometric redshifts for all the galaxies, which range from z$_{phot}$=0.05 
to 3. By dividing the sample according to the fitted templates, in \citealt{Ramos09} we proposed 
an evolutionary sequence, 
similar to the one suggested for early-type galaxies by \citet{Schawinski07}:
an intense period of star formation would be quenched by AGN feedback as the BH accretes enough mass, 
competing for the cold gas reservoir and heating it, and becoming dominant. The AGN-phase would 
then continue through lower ionization phases, ending as spiral or elliptical galaxies hosting low-luminosity AGN.

This paper constitutes a spectroscopic follow-up of a representative subset 
of the AGN sample studied in \citealt{Ramos09}. These AGN were selected in the X-rays 
and all of them have MIR counterparts (i.e. detection in the 3.6, 4.5, 5.8, 8, and 24 \micron~bands
of the Spitzer Space Telescope). See Section \ref{sample} for further 
details in the sample selection. Here we present NIR
spectroscopic observations for 28 of these AGN, which have spectroscopic redshifts in the range z=[0.27, 1.28], 
and a median redshift of z=0.76. 
These objects are representative of the whole sample in terms of redshift, magnitude, and 
SED types (see Section \ref{sample}), including 13 AGN-dominated and 15 host 
galaxy-dominated objects. The main goals of this work are first, to classify the galaxies spectroscopically
to check the reliability of the SED classification done in \citealt{Ramos09}, 
and second, obtain SFRs using the H$\alpha$ emission. We will compare these SFRs with those obtained using 
Spitzer 24 \micron~observed fluxes.  
%The EGS ($\alpha$ = 14$^{h}$ 17$^{m}$, $\delta$ = +52$^{o}$ 30') enlarges the Hubble Space
%Telescope Groth-Westphal strip \citep{Groth94} up to 2$^{o}$x15', having the advantage of
%being a low extinction area in the northern sky, with low galactic and zodiacal infrared emission, and
%good schedulability by space observatories. For these reasons, there is a vast amount of public data at
%different wavelength ranges that only require to be compiled and cross-correlated in a consistent way.
%The overall majority of the observational work in the EGS
%have been coordinated by the AEGIS proyect\footnote{The AEGIS project is a collaborative effort to obtain both 
%deep imaging covering all major wavebands from X-ray to radio and optical spectroscopy over a large area 
%of sky. http://aegis.ucolick.org/index.html} \citep{Davis07}. 
Throughout this paper  we assume a cosmology with H$_0 = 75~km~s^{-1}~Mpc^{-1}$, $\Omega_{m}$=0.27, and 
$\Omega_{\Lambda}$=0.73.

\section{Sample and photometric data}
\label{sample}

The AGN sample studied in \citealt{Ramos09} was originally selected by \citet{Barmby06}
in the X-rays, using Chandra data from \citet{Nandra05} and XMM-Newton data from \citet{Waskett04}.
\citet{Barmby06} considered only the 152 X-ray sources 
lying within the boundaries of the Spitzer observations with a limiting full-band flux (0.5-10 keV) 
of 2$\times10^{-15} erg s^{-1} cm^{-2}$ in the
case of the XMM-Newton data, and of 3.5$\times10^{-16} erg s^{-1} cm^{-2}$ for the Chandra data. 
At the flux limits of these X-ray surveys, most of the sources are expected to be AGN and 
have log(f$_X/f_{opt}) >$ -1, indicating that they are not quiescent (i.e. non-active) galaxies.
Finally, they selected the 138 objects with secure detections in the four Spitzer/IRAC bands 
(3.6, 4.5, 5.8, and 8 \micron) and Spizer/MIPS 24 \micron~band. 

In addition to the X-ray and MIR data, in \citealt{Ramos09} we used optical and NIR archival data from the AEGIS 
proyect\footnote{The AEGIS project is a collaborative effort to obtain both 
deep imaging covering all major wavebands from X-ray to radio and optical spectroscopy over a large area 
of sky. http://aegis.ucolick.org/index.html} \citep{Davis07} to increase the coverage of the AGN SEDs. 
All the fluxes employed in \citealt{Ramos09} were directly retrieved from 
the {\it Rainbow Cosmological Surveys database}\footnote{https://rainbowx.fis.ucm.es/Rainbow$_{-}$Database},    
which is a compilation of photometric and spectroscopic data, jointly with value-added products such as photometric 
redshifts and synthetic rest-frame magnitudes, for several deep cosmological fields 
\citep{Perez08b,Barro09,Barro11}.
Of the 138 sources in the \citet{Barmby06} sample, we discarded 42 galaxies that showed multiple 
detections in the ground-based images (optical and NIR) to 
avoid source confusion in the Spitzer MIR fluxes.

To classify the observed SEDs and estimate photometric redshifts, 
we combined optical data from the Canada-France-Hawaii Telescope Legacy Survey 
(CFHTLS; u,g,r,i,z) T0003 worldwide release \citep{Gwyn12}; NIR fluxes (J and K$_S$) from the 
version 3.3 of the Palomar-WIRC K-selected catalog \citep{Bundy06}, and Spitzer MIR data 
(IRAC 3.6, 4.5, 5.8, 8 \micron~and MIPS 24 \micron) from \citet{Barmby06}. We fitted these SEDs 
with the library of starburst, AGN, and galaxy templates from \citet{Polletta07} using the photometric redshift code
HyperZ \citep{Bolzonella00}. The templates span the wavelength range 0.1--1000 \micron. 
See \citealt{Ramos09} for a detailed description of the data, 
photometric redshift calculations, and SED classification. 

The galaxies were then classified in five main categories in terms of the template used to fit
their SEDs\footnote{http://www.iasf-milano.inaf.it/$\sim$ polletta/templates/swire$_-$templates.html}:
\begin{enumerate}

\item {\it Starburst-dominated}. Includes several templates of starburst galaxies (e.g. M82-like) 
and ultraluminous infrared galaxies (ULIRGs) with starburst (e.g. Arp 220-like). 

\item {\it Starburst-contaminated}. Comprises three different starburst-composite SEDs: one 
starburst/Seyfert 1 and two starburst/Seyfert 2.

\item {\it Type-1}. Includes three Type-1 quasar (QSO) SEDs.

\item {\it Type-2}. Consists of two Type-2 QSOs, a Seyfert 2 and a Seyfert 1.8 SED. 

\item {\it Normal galaxy}. Includes the SEDs of three elliptical galaxies and seven spirals of different
types.

\end{enumerate}

Our main goal in \citealt{Ramos09} 
was to classify the galaxies into these five groups and study their properties.
It is worth clarifying that all of the sources are, in principle, AGN, on the basis of their X-ray and MIR emission, 
but only those included in the {\it Type-1}, {\it Type-2} and {\it Starburst-contaminated (SB-cont)}
groups have AGN-dominated SEDs.
On the other hand, those included in the {\it Starburst-dominated (SB-dom)} group have their SEDs 
dominated by starburst emission from the optical to the MIR. Finally, the 
{\it Normal galaxy (NG)} are weak AGN embedded in an otherwise elliptical or spiral galaxy
emission. 
%Finally, in the case of the {\it Starburst-contaminated} group {\it (SB-cont)}, the emission of either the starburst 
%or the AGN dominate depending on the wavelength we look at (see Figure 1 in \citealt{Ramos09}). 

It is possible, however, that a small fraction of the sample are star-forming galaxies 
emitting in the X-rays and MIR (see e.g. \citealt{Pereira11,Ranalli12}).
In order to confirm the dominance/presence of the AGN over the star formation/host galaxy emission 
and vice versa, and to estimate 
SFRs from the H$\alpha$ emission, we obtained
NIR spectroscopic data for a subsample of 28 galaxies ($\sim$30\% of the total sample), which are representative 
of the five groups described above (see Table \ref{photoz}). In the following, we will refer to these 28 
sources as AGN, although one of the goals of this work is to confirm the presence of nuclear activity.

\begin{figure*}
\centering
{\par
\includegraphics[width=8cm]{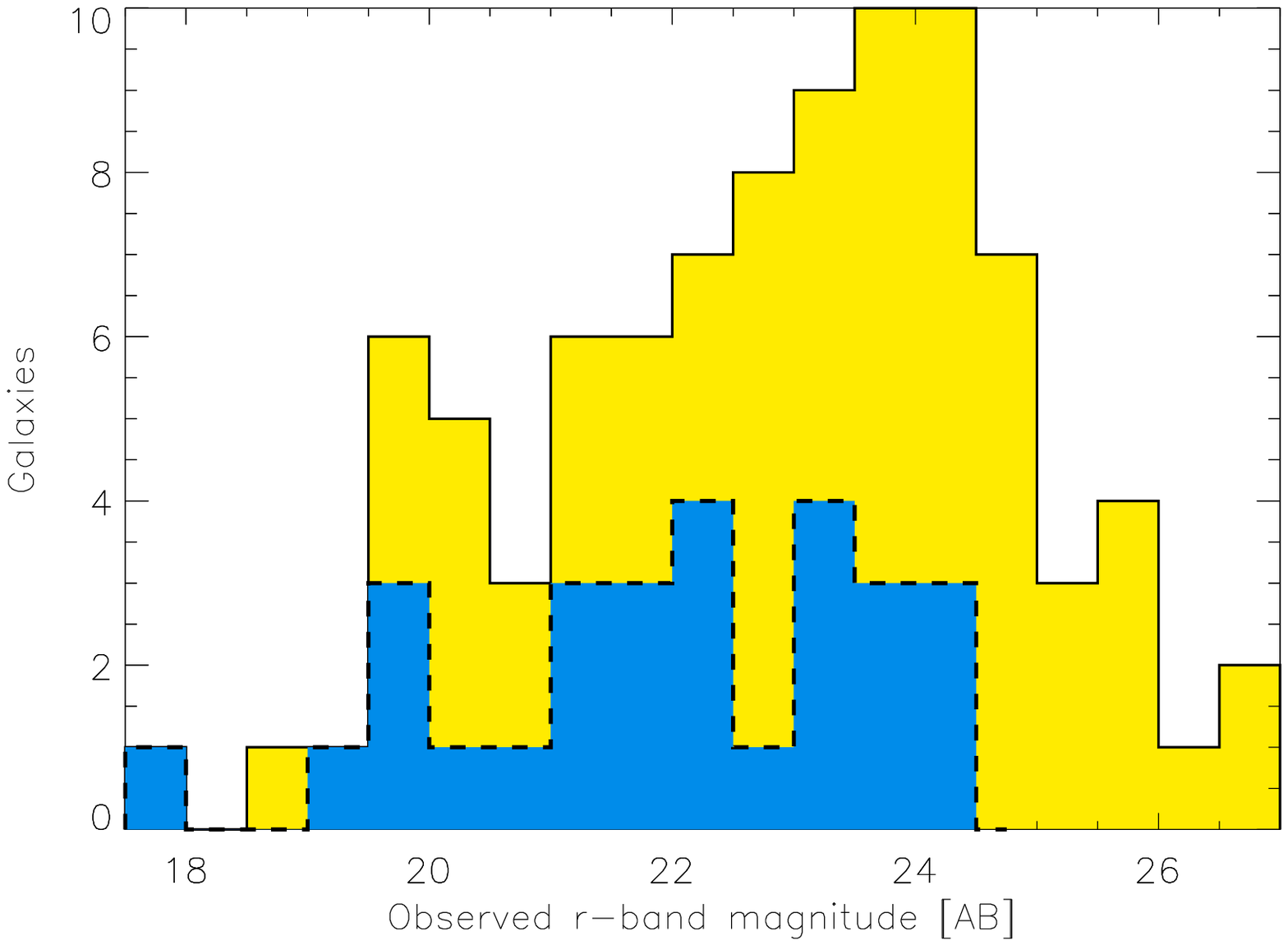}
\includegraphics[width=8cm]{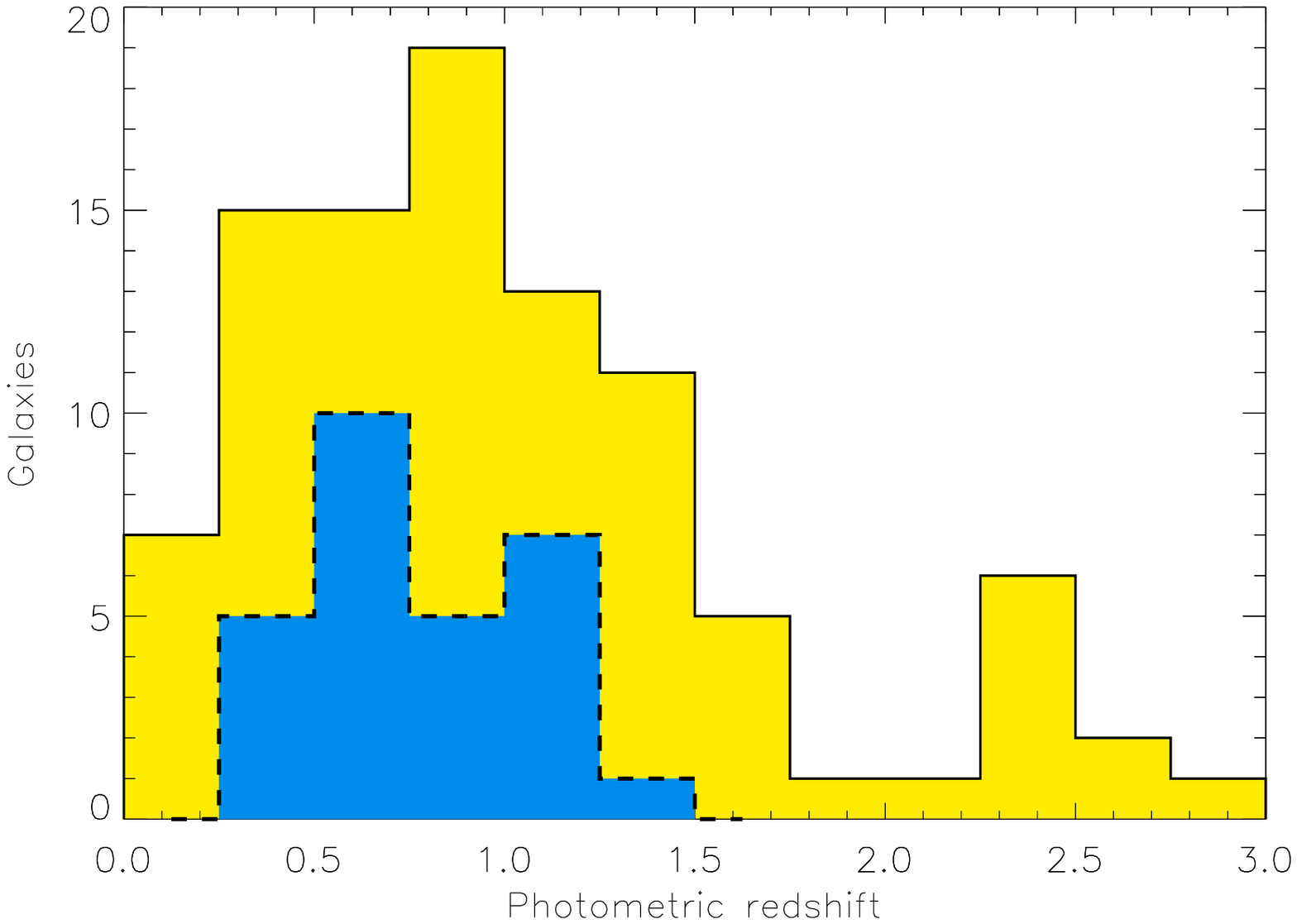}\par}
\caption{Left panel: observed r-band magnitudes for the whole AGN sample studied in \citealt{Ramos09} 
(yellow and continuous line histogram) and for the subsample of 28 galaxies observed with LIRIS (blue and dashed 
line histogram). Right panel: same as in the left panel, but for the photometric redshift distributions.}
\label{doherty}
\end{figure*}

In order to discard any possible bias in 
redshift and/or magnitude, in Figure \ref{doherty} we show histograms for the observed subsample and 
the whole \citealt{Ramos09} sample. According to the Kolmogorov-Smirnov (KS) test, there is no
significant difference between the two samples in terms of observed magnitude and redshift 
at the 2$\sigma$ level.

\begin{figure*}
\centering
\includegraphics[width=12cm]{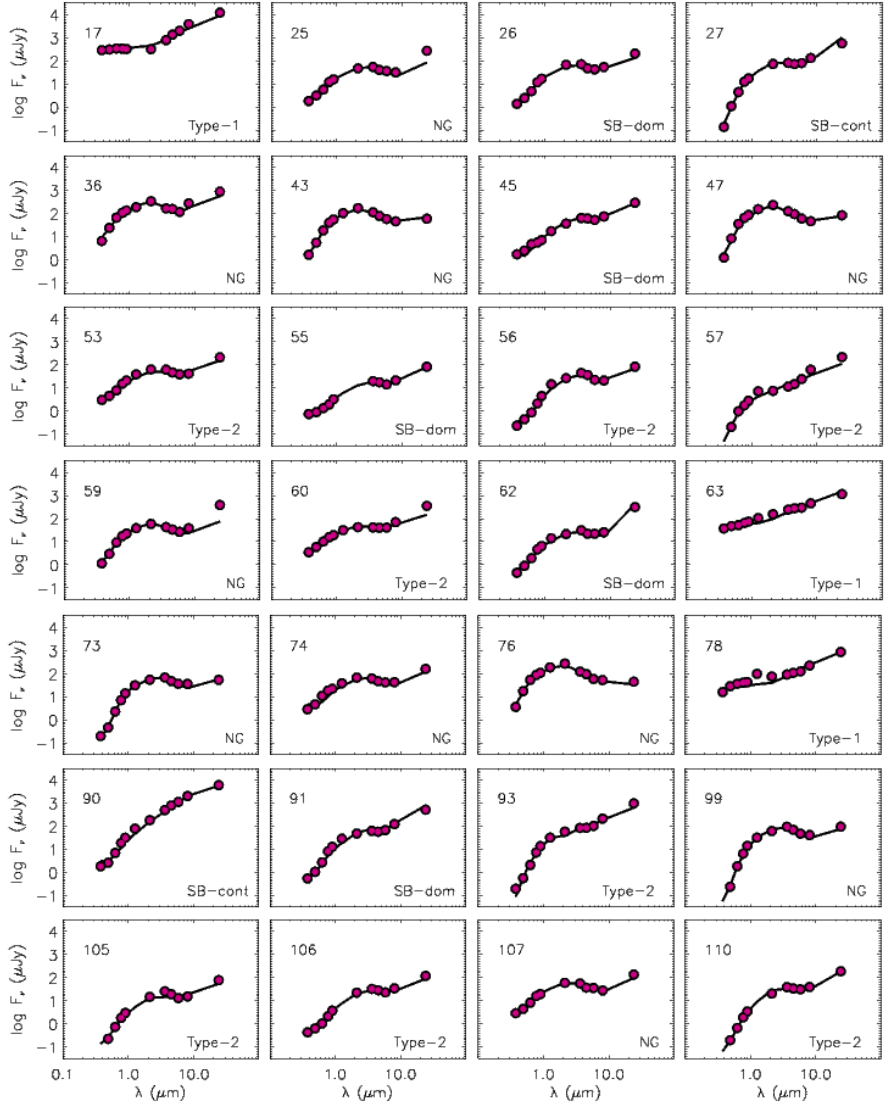}
\caption{Observed SEDs (dots) of the subsample of 28 AGN observed with LIRIS, fitted with the
AGN, starburst, and composite templates from \citet{Polletta07}, using HyperZ. 
The SED group and galaxy ID from \citealt{Ramos09} are indicated in each panel.}
\label{seds}
\end{figure*}

In Figure \ref{seds} we show the observed galaxy
SEDs and the templates fitted in \citealt{Ramos09} to classify them 
in the previously mentioned groups, as well as to estimate photometric redshifts.  
The fits are the same presented in \citealt{Ramos09}, with the exception of those of the 
galaxies G17, G57, G78, and G105. We repeated these four fits because the photometric 
redshifts were not compatible with the spectroscopic redshifts derived from the NIR
spectra presented here ($\mid z_{NIR}-z_{phot}\mid > 0.2$). We performed the 
fits using HyperZ and restricting the input redshift range to $z_{NIR}\pm{0.08}$, 
which is the maximum difference between $z_{NIR}$ and $z_{phot}$ found for the rest of the sample
observed with LIRIS. This sample includes 5 {\it SB-dom}, 2 {\it SB-cont}, 
3 {\it Type-1}, 8 {\it Type-2}, and 10 {\it NG}, which 
%We chose the NIR range first, to avoid dust obscuration, which can easily mask the AGN contribution, 
%and second, because considering the median redshift 
%of the total sample (z=0.9), the H$\alpha$ line lies in the J-band. 
were selected to cover the five SED groups and to maximize the number of targets 
in the LIRIS multi-slit masks (see Section \ref{nir}).

\begin{figure*}
\centering
\includegraphics[width=15cm]{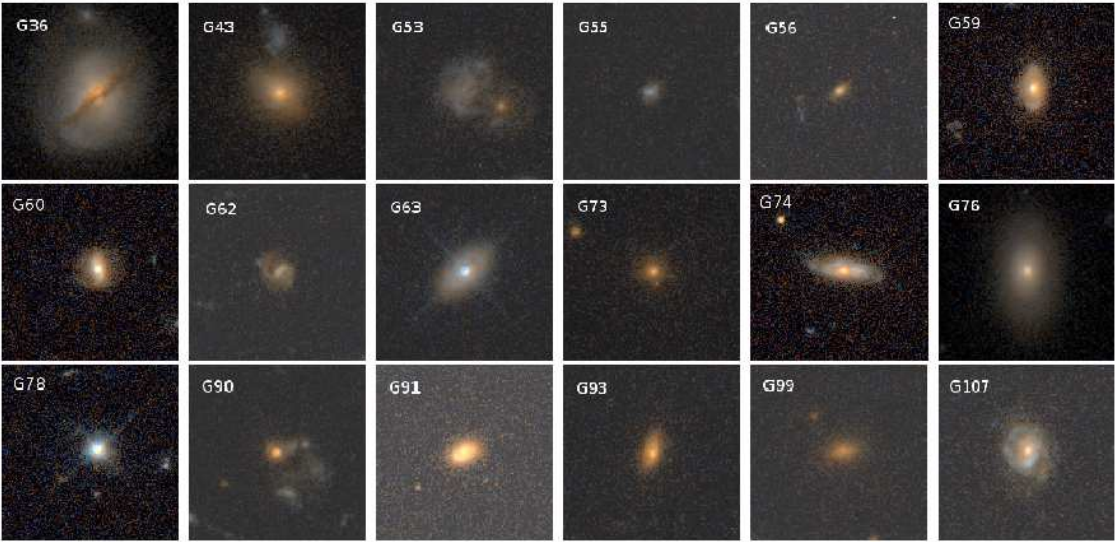}
\caption{ACS/HST color-composite images of the 18 AGN in our LIRIS representative sample 
lying in the AEGIS ACS fields. Images are 7 arcsec size and were downloaded from the AEGIS website 
({\it http://tkserver.keck.hawaii.edu/egs/egs.php}). North is up and East is left.}
\label{acs}
\end{figure*}

Color-composite images obtained with
the Advanced Camera for Surveys (ACS) on the Hubble Space Telescope (HST) for the 18 galaxies with 
data available in the AEGIS website\footnote{http://tkserver.keck.hawaii.edu/egs/} are shown in Figure \ref{acs}.
The {\it Type-1} G63 and G78 show blue colors and strong point sources, while galaxies classified 
as either {\it Type-2} (e.g. G53) or {\it SB-dom} (e.g. G55 and G62) show obscured
nuclei. Examples of spiral galaxies are G74 and G107, which were fitted with spiral templates in 
\citealt{Ramos09}. Finally, the morphology of G36 resembles that of an elliptical galaxy with a strong dust lane
crossing the nucleus.

\begin{table*}
\centering
\begin{tabular}{lcccccccc}
\hline
\hline
ID & ID DEEP2 & RA ($\degr$) & Dec ($\degr$) & J (AB)  & Log (L$_X$) & z$_{spec}$  &
z$_{phot}$ &  Group \\ 
\hline
G17  & 11039094 &  214.1768  & 52.3034  & [17.41] & 44.53     & 1.284 (a)  &    1.25 &   3     \\
G25  & 11038472 &  214.2065  & 52.2815  & [20.40] & 42.97     & 0.761 (4)  &    0.75 &   5     \\
G26  & 11038492 &  214.2079  & 52.3025  & [20.27] & 43.37     & 0.808 (4)  &    0.73 &   1     \\
G27  & 11038266 &  214.2104  & 52.2763  & [20.00] & 43.45     & 0.683 (4)  &    0.60 &   2     \\
\hline
G36  & 12004450 &  214.2675  & 52.4149  & 18.22   & 42.52     & 0.281 (4)  &   0.25  &   5     \\
G43  & 12008608 &  214.2870  & 52.4525  & 18.87   & 42.08     & 0.532 (4)  &   0.47  &   5     \\
G45  & 12008576 &  214.2940  & 52.4747  & 20.81   & 43.59     &   -        &   1.25  &   1     \\
G47  & 12004491 &  214.2961  & 52.4280  & 18.44   &$\leq$41.79& 0.418 (b)  &   0.34  &   5     \\
G53  & 12004467 &  214.3134  & 52.4474  & 19.94   & 43.13     & 0.723 (4)  &   0.67  &   4     \\
G55  & 12008051 &  214.3290  & 52.4623  & [21.93] &$\leq$43.03& 1.211 (3)  &   1.23  &   1     \\
G56  & 12008091 &  214.3303  & 52.4655  & 21.01   & 42.92     & 1.208 (3)  &   1.19  &   4     \\  
G57  & 12004011 &  214.3335  & 52.4168  & 21.77   & 42.71     &   -	   &   0.70  &   4     \\
\hline
G59  & 12012474 &  214.3456  & 52.5288  & 19.93   & 42.06     &  0.465 (4) &   0.46  &   5     \\
G60  & 12012471 &  214.3475  & 52.5316  & 20.14   & 42.68     &  0.484 (4) &   0.50  &   4     \\ 
G62  & 12012431 &  214.3510  & 52.5416  & 21.02   &$\leq$42.55&  0.902 (4) &   0.83  &   1     \\ 
G63  & 12008225 &  214.3525  & 52.5069  & 18.84   & 43.31     &  0.482 (4) &   0.54  &   3     \\ 
G73  & 12012467 &  214.3859  & 52.5342  & 20.18   & 43.61     &  0.986 (4) &   0.91  &   5     \\
G74  & 12012543 &  214.3909  & 52.5637  & 19.95   & 42.70     &  0.551 (4) &   0.52  &   5     \\
G76  & 12012534 &  214.3932  & 52.5186  & 18.22   & 41.48     &  0.271 (4) &   0.29  &   5     \\
G78  & 12008222 &  214.3998  & 52.5083  & 18.93   & 44.66     &  0.985 (c) &   1.04  &   3     \\
\hline
G90  & 12007954 &  214.4244  & 52.4732  & 19.16   & 44.40     &  1.148 (4) &  1.15   &   2     \\
G91  & 12007926 &  214.4393  & 52.4976  & 20.25   & 43.33     &  0.873 (4) &  0.87   &   1     \\
G93  & 12007878 &  214.4415  & 52.5091  & 20.14   & 43.89     &  0.985 (3) &  0.97   &   4     \\
G99  & 12007962 &  214.4550  & 52.4676  & 20.13   & 42.88     &  0.996 (4) &  1.00   &   5     \\
G105 & \dots    &  214.4657  & 52.5129  & [22.10] & 41.87     &	-          &  0.57   &   4     \\
G106 & 12007949 &  214.4684  & 52.4814  & [21.69] & 43.22     &	-          &  1.00   &   4     \\
G107 & 12007896 &  214.4707  & 52.4775  & [20.10] &$\leq$42.04&  0.671 (3) &  0.60   &   5     \\
G110 & 12012132 &  214.4760  & 52.5232  & [21.51] & 42.56     &	-          &  0.65   &   4     \\
\hline
\end{tabular}
\caption{Columns 1 and 2 list the ID from \citet{Barmby06} and from the DEEP2 database. Columns 3, 4, and 5 give the 
IRAC 3.6~\micron~ right ascension and declination and the J mag from the Palomar-WIRC K-selected catalog. For the galaxies lacking of 
Palomar J mags, values from the SED fits are given between brackets instead. Column 6 lists the observed hard X-ray luminosities 
from \citet{Nandra05} and \citet{Waskett04} in erg~s$^{-1}$.
Column 7 lists the spectroscopic redshift from DEEP2 when available (reliability is given between brackets: 3=robust and 4=very robust) and 
from other literature sources for the galaxies G17, G47, and G78. Columns 8 and 9 list the photometric redshift and the 
SED classification from \citealt{Ramos09} (1 = SB-dom; 2 = SB-cont; 3 = Type-1; 4 = Type-2; 5 = NG).
Refs: (a) \citet{Schneider05}; (b) \citet{Steidel03}; (c) \citet{Lilly95}.}
\label{photoz}
\end{table*}

%The five groups can also be segregated using the 24 \micron to optical r-band flux ratio versus either the (r-z) or 
%(r-3.6 \micron)colors (see Figure 6 in \citealt{Ramos09}). {\it Starburst-dominated AGNs} and {\it Starburst-contaminated AGNs} 
%are displaced toward the highest values of the MIR-to-optical ratio and display the reddest colors. 
%{\it Type-1 AGNs} and {\it Type-2 AGNs} are located at intermediate values, and the {\it Normal galaxy hosting AGNs} have 
%the lowest values of the flux ratio and the bluest colors.

%\begin{figure*}
%\centering
%{\par
%\includegraphics[width=2.5cm]{../all/images/G36_acs.ps}
%%\includegraphics[width=2.5cm]{../all/images/G43_acs.ps}
%\includegraphics[width=2.5cm]{../all/images/G53_acs.ps}
%\includegraphics[width=2.5cm]{../all/images/G55_acs.ps}
%\includegraphics[width=2.5cm]{../all/images/G56_acs.ps}
%\includegraphics[width=2.5cm]{../all/images/G59_acs.ps}
%\includegraphics[width=2.5cm]{../all/images/G60_acs.ps}
%\includegraphics[width=2.5cm]{../all/images/G62_acs.ps}
%\includegraphics[width=2.5cm]{../all/images/G63_acs.ps}
%\includegraphics[width=2.5cm]{../all/images/G73_acs.ps}
%\includegraphics[width=2.5cm]{../all/images/G74_acs.ps}
%\includegraphics[width=2.5cm]{../all/images/G76_acs.ps}
%\includegraphics[width=2.5cm]{../all/images/G78_acs.ps}
%\includegraphics[width=2.5cm]{../all/images/G90_acs.ps}
%\includegraphics[width=2.5cm]{../all/images/G91_acs.ps}
%\includegraphics[width=2.5cm]{../all/images/G93_acs.ps}
%\includegraphics[width=2.5cm]{../all/images/G99_acs.ps}
%\includegraphics[width=2.5cm]{../all/images/G107_acs.ps}\par}
%\caption{\footnotesize{ACS/HST color-composite images of the 18 AGN in our LIRIS representative sample 
%lying in the AEGIS ACS fields. Images are 7 arcsec size and were downloaded from the AEGIS website 
%({\it http://tkserver.keck.hawaii.edu/egs/egs.php}). North is up and East is left.}
%\label{acs}}
%\end{figure*}

%ESTA FIGURA ESTA SE GENERA EN EL DIRECTORIO /local/cra/cristina/iac/apr2011/groth/qso_star/UGRIZJKIRACM/
%leedatos_aegis_apr2011.pro

\section{Observations and data reduction}
\label{obs}

\subsection{NIR spectroscopy}
\label{nir}

NIR spectroscopic observations of the subsample of 28 galaxies selected from \citealt{Ramos09} 
were obtained from 2008 March to 2009 May 
using the multi-object spectroscopic (MOS) mode of the NIR camera/spectrometer LIRIS \citep{Manchado04}.
LIRIS is attached to the Cassegrain focus of the 4.2 m WHT and 
it is equipped with a Rockwell Hawaii 1024 x 1024 HgCdTe array detector. The spatial scale is
0.25\arcsec~pixel$^{-1}$. 

Four masks were designed to observe a representative subset of 
the five groups described in Section \ref{sample}.
The galaxies were selected by their RA and Dec., to maximize the number of targets per mask. 
Details of each mask and the journal of observations are reported in Table \ref{log}. 
The chosen slit-width was 0.85\arcsec, with the lengths varying between 8.5\arcsec~and 12\arcsec, allowing enough space 
for nodding while avoiding overlap of the spectra.
We used the low-resolution grism ZJ, which covers the range 0.8--1.4 \micron, providing 
a spectral resolution of $\sim$500 km~s$^{-1}$ with the 0.85\arcsec~slits. 
The spectral range varies depending on the position of the slits on the mask along the spectral direction. 
Thus, a slit in the center of the mask will provide a spectrum in the nominal spectral range (0.8--1.4 \micron), 
whereas spectra obtained with slits closer to the edges of the mask will be shifted either bluewards or redwards.

In addition to the science targets slitlets (see details in Table \ref{log}), each mask 
contained an extra-slit, designed to simultaneously obtain the spectrum of a star of $\sim$16-17 mag in the J-band (AB).
These stars, which are brighter than the science targets (see Table \ref{photoz}), allowed, on the one hand, to ensure 
that the masks were  well-centred during the observation (the majority of the targets are too faint to be detected in  
single exposures), 
and on the other hand, to calculate the corresponding correction once the spectra were flux-calibrated. 

\begin{table*}
\centering
\begin{tabular}{lccccccc}
\hline
\hline
Mask & Slits & Position & P.A. & Obs. Date &\multicolumn{1}{c}{Exposure Time}  &\multicolumn{1}{c}{Airmass} & Seeing \\
\hline
EGS1 & 8  &  214.375, +52.535 &   5$\degr$ & 2008 Mar 28     & 4x6x600 s    & 1.09-1.35  & $\sim$1\arcsec	  \\
EGS2 & 8  &  214.293, +52.446 &   9$\degr$ & 2009 May 9      & 4x6x600 s    & 1.09-1.28  & $\sim$0.6\arcsec	  \\
EGS3 & 8  &  214.458, +52.498 &  26$\degr$ & 2009 May 8      & 4x6x600 s    & 1.09-1.41  & $\sim$0.7-0.9\arcsec    \\
EGS5 & 4  &  214.182, +52.283 &  26$\degr$ & 2009 May 8,9    & 2x6x600 s    & 1.55-1.98  & $\sim$0.6-0.7\arcsec    \\
\hline  					 					
\end{tabular}
\caption{Summary of the NIR multi-object spectroscopic observations}
\label{log}
\end{table*}

We performed the observations following an ABCABC telescope-nodding pattern, placing the source in three
positions along the slit, using an offset of 3\arcsec~around the central position of the pattern (B). 
Individual frames of 600 s in each nodding position were taken (see Table \ref{log}), amounting a total integration 
time of 3600 s per ABCABC pattern, and a total of four hours per mask. Re-centering of the masks 
was done every hour, since the relative position between the star used for guiding and the science targets 
might have changed due to differential atmospheric refraction.   
The wavelength calibration was provided by
observation of the argon lamp available in the calibration unit at the A\&G box of the telescope. To
obtain the telluric correction and the flux calibration for each galaxy, the nearby G0 stars HD136674 and 
HD115269 were observed before and after the science targets. Spectra of the two stars were obtained through 
three slitlets covering the whole wavelength coverage.

We reduced the data following stardard procedures for NIR spectroscopy, using the 
{\it lirisdr}\footnote{http://www.iac.es/project/LIRIS} dedicated software within the IRAF\footnote{IRAF
is distributed by the National Optical Astronomy Observatory, which is operated by the Association of
Universities for the Research in Astronomy, Inc., under cooperative agreement with the National Science
Foundation (http://iraf.noao.edu/).} enviroment. For the MOS mode, the  available routines use 
a-priori mask design information to trace the slitlets positions and limits in a uniformly illuminated
frame. We did not perform flat-field correction after checking that, at least in the case of our observations, 
it only introduced additional noise. 
We subtracted the mean of alternate pairs (A+C)/2 of two-dimensional spectra from that taken in 
the B position of the nodding pattern to remove the sky background. 
The resulting frames were wavelength-calibrated before registering and co-adding all frames 
to provide the final spectra. 

We then extracted the individual spectra for each galaxy and reference star, using 
an aperture of 1\arcsec, matching the maximum value of the seeing during the observations (see Table \ref{log}). 
We finally divided the extracted spectra by the corresponding G0 star spectrum to remove telluric
contamination. We used a modified version of {\it Xtellcor} \citep{Vacca03} in this step, which
provides both the telluric correction and the flux calibration. 

\begin{figure*}
\centering
{\par
\includegraphics[width=12cm]{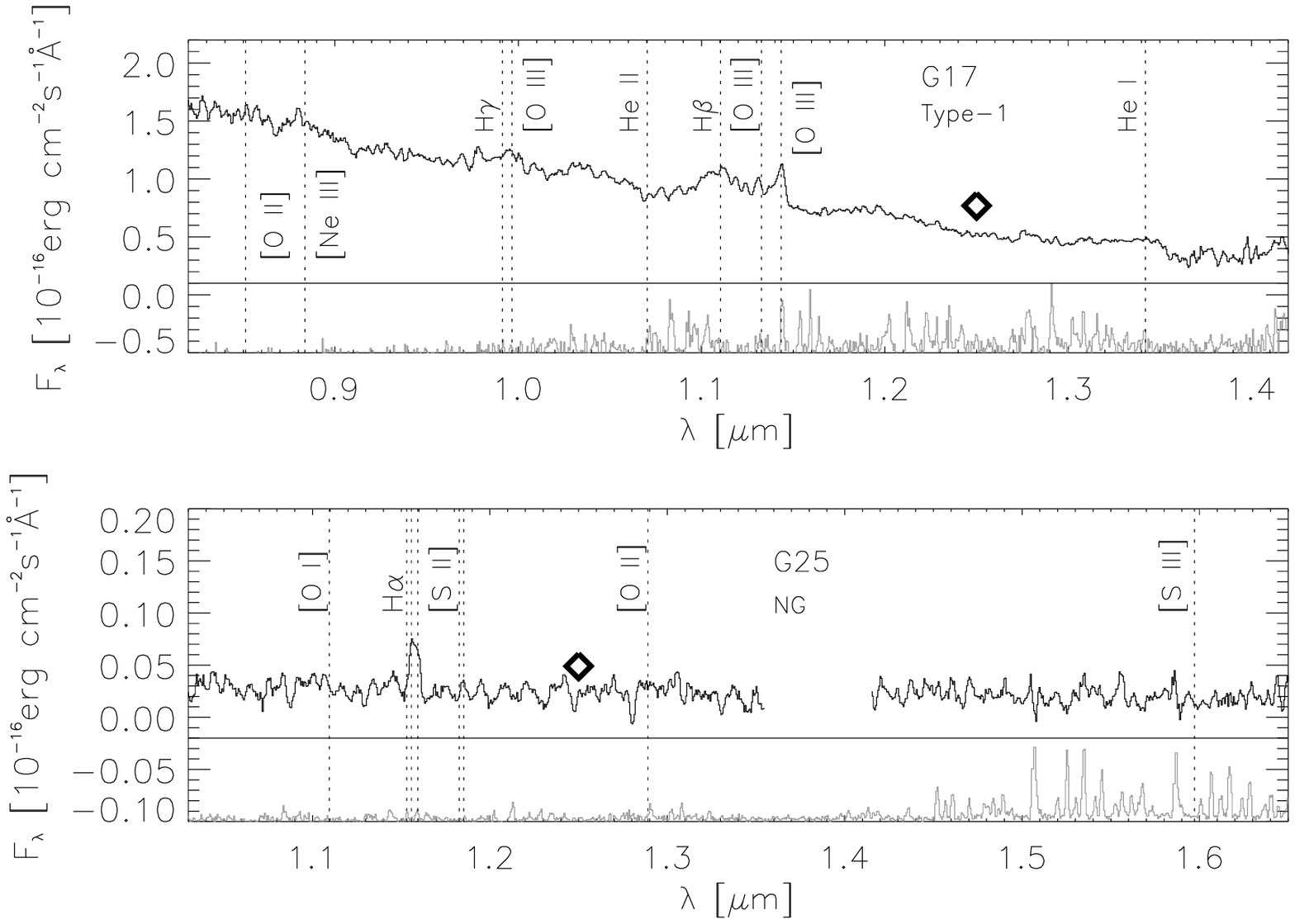}
\includegraphics[width=12cm]{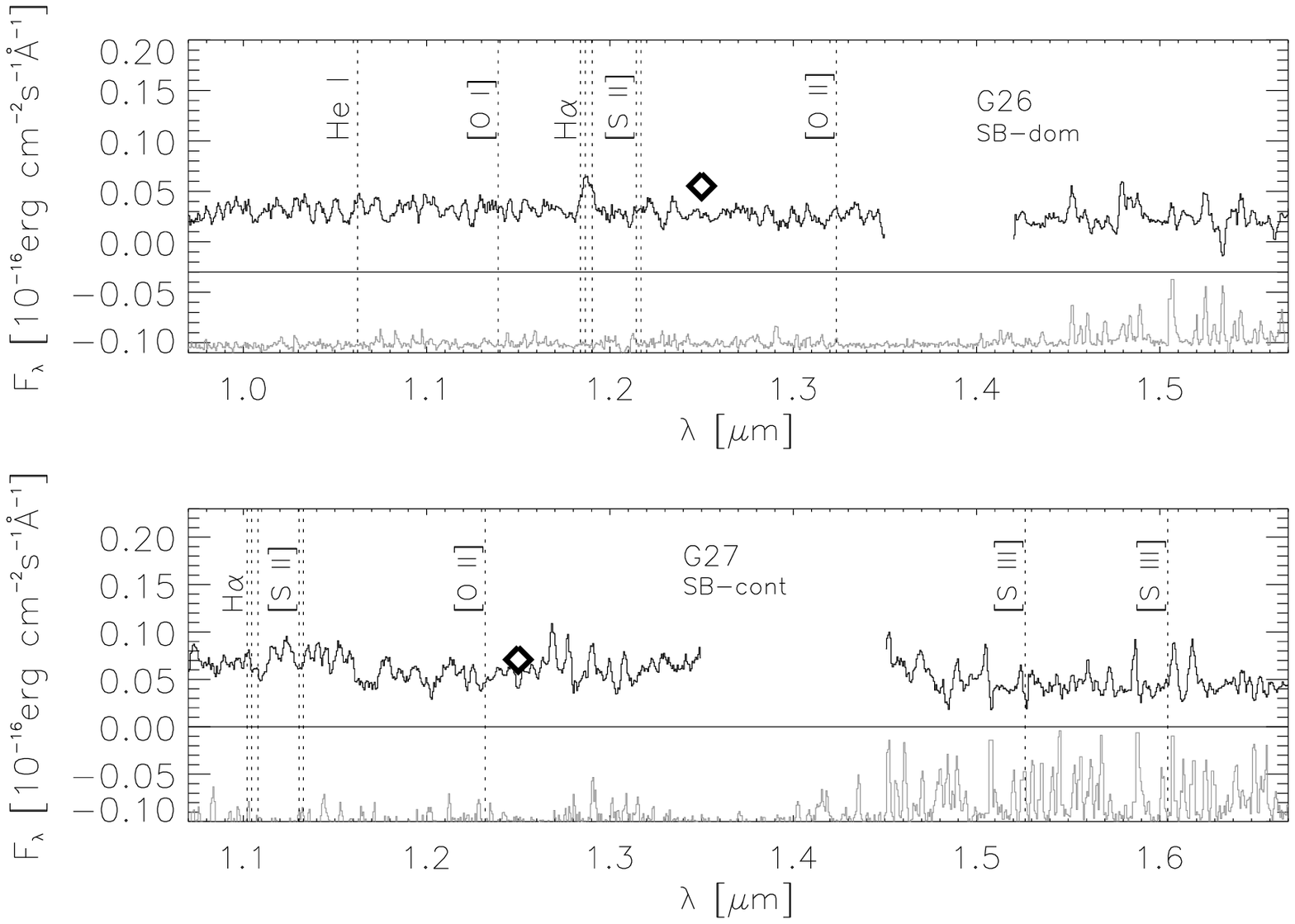}\par}
\caption{Observed NIR LIRIS spectra of the galaxies G17, G25, G26, and G27. Typical AGN emission lines are
labelled. The H$\alpha$ labels correspond to H$\alpha$+2[N II].
The SED classification from \citealt{Ramos09} is indicated at the top of each panel, and 
an scaled sky spectrum for each galaxy is plotted at the bottom. Observed J-band fluxes from the magnitudes
reported in Table \ref{photoz} are represented with a cross for comparison. Open diamonds correspond to 
the J-band fluxes from the SED fits shown in Figure \ref{seds}.}
\label{groth5}
\end{figure*}

\begin{figure*}
\centering
{\par
\includegraphics[width=12cm]{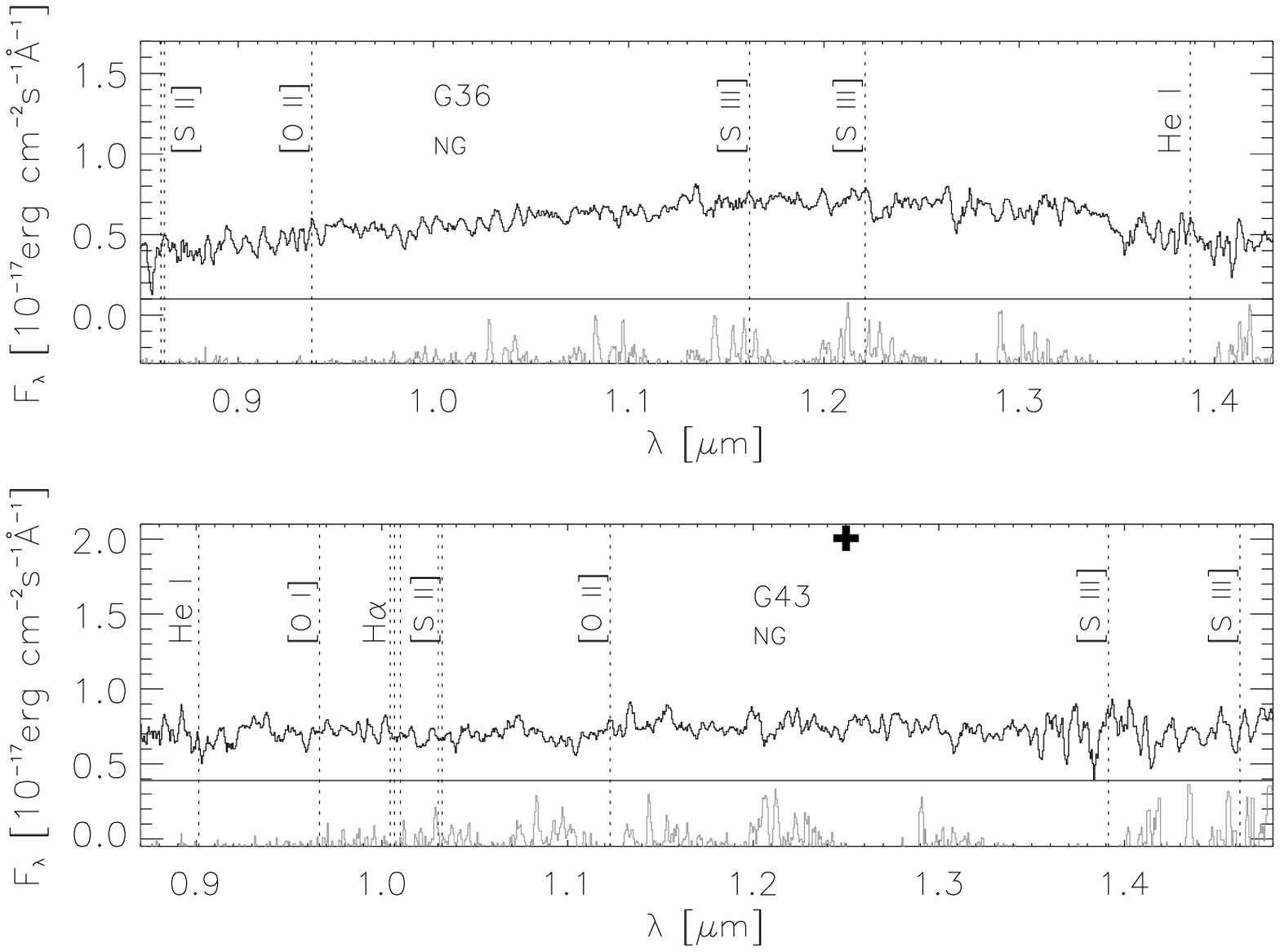}
\includegraphics[width=12cm]{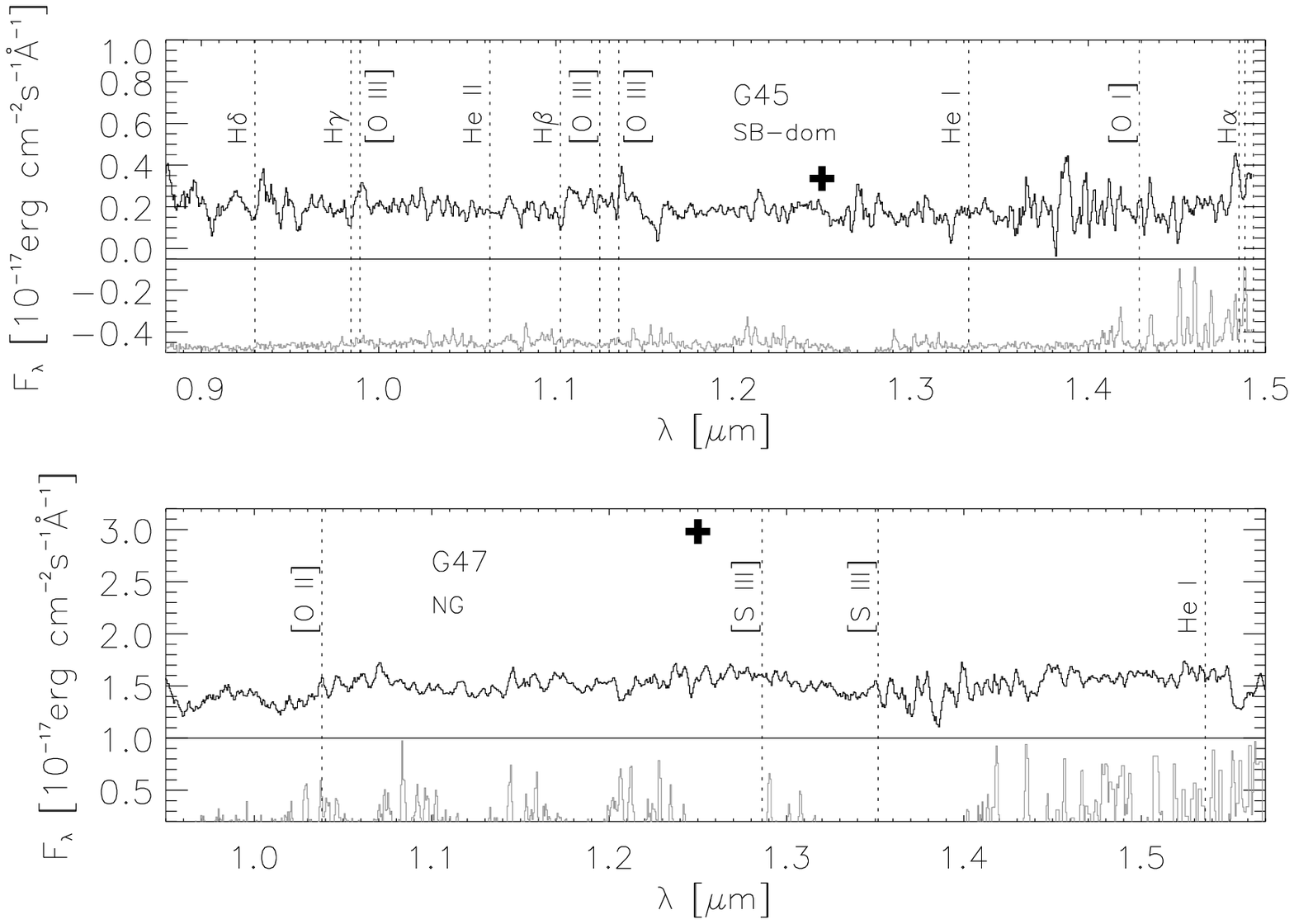}\par}
\caption{Same as in Figure \ref{groth5} but for the galaxies G36, G43, G45, and G47. The J-band 
flux of the galaxy G36 is out of scale (J flux = 3.65$\times 10^{-17}~erg~cm^{-2}~s^{-1}~\AA^{-1}$).}
\label{groth2a}
\end{figure*}

\begin{figure*}
\centering
{\par
\includegraphics[width=12cm]{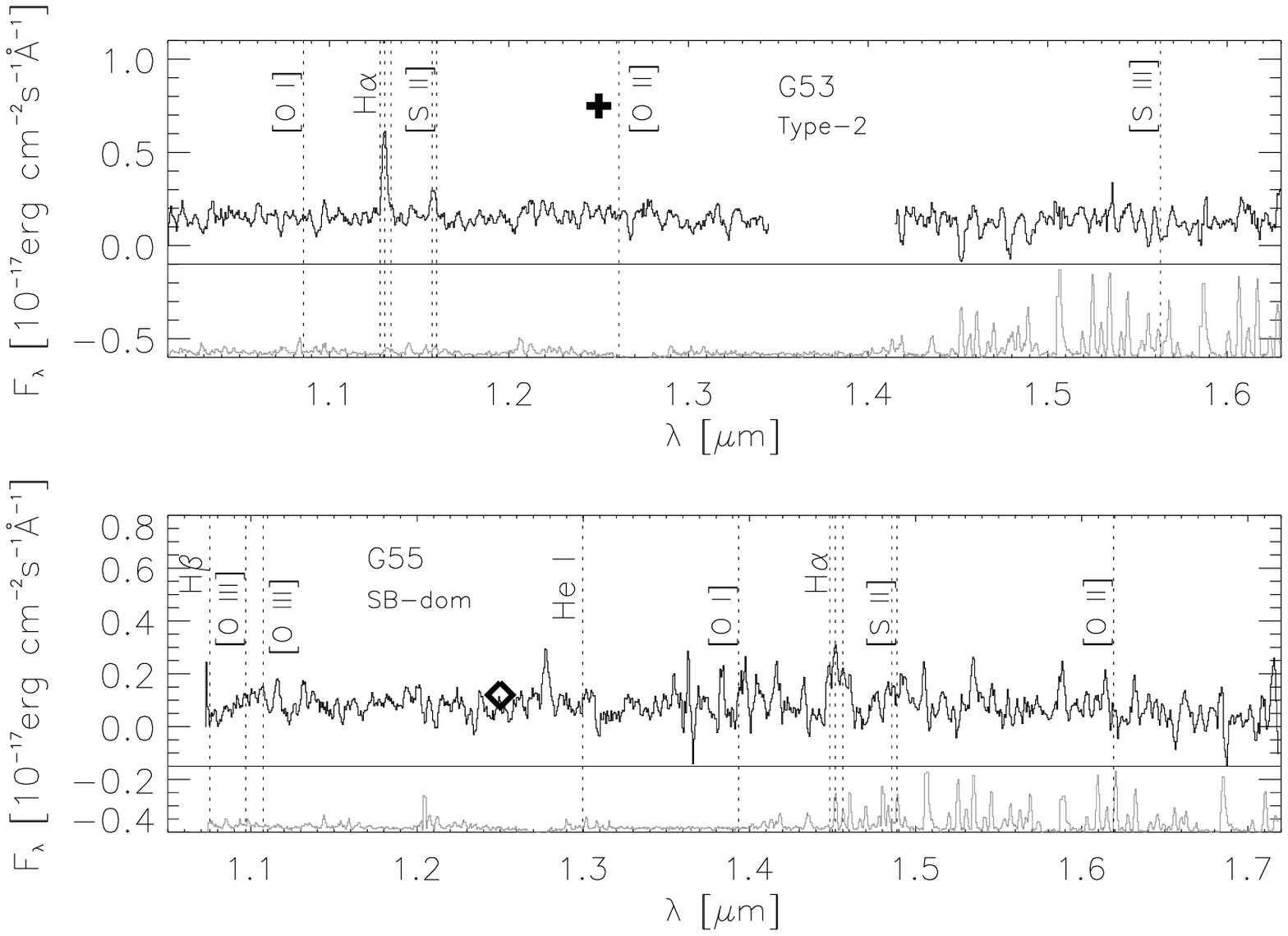}
\includegraphics[width=12cm]{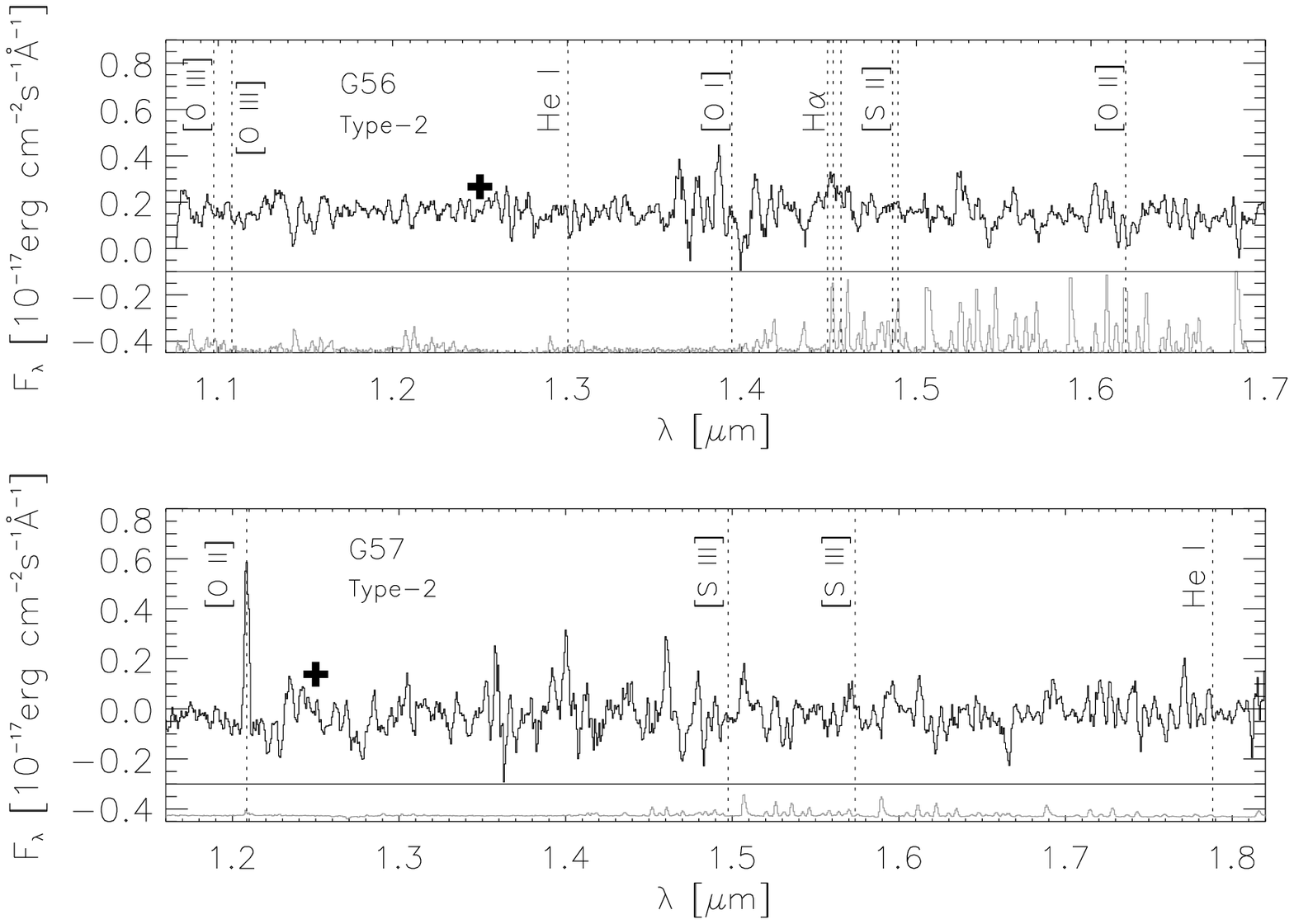}\par}
\caption{Same as in Figure \ref{groth5} but for the galaxies G53, G55, G56, and G57.}
\label{groth2b}
\end{figure*}

\begin{figure*}
\centering
{\par
\includegraphics[width=12cm]{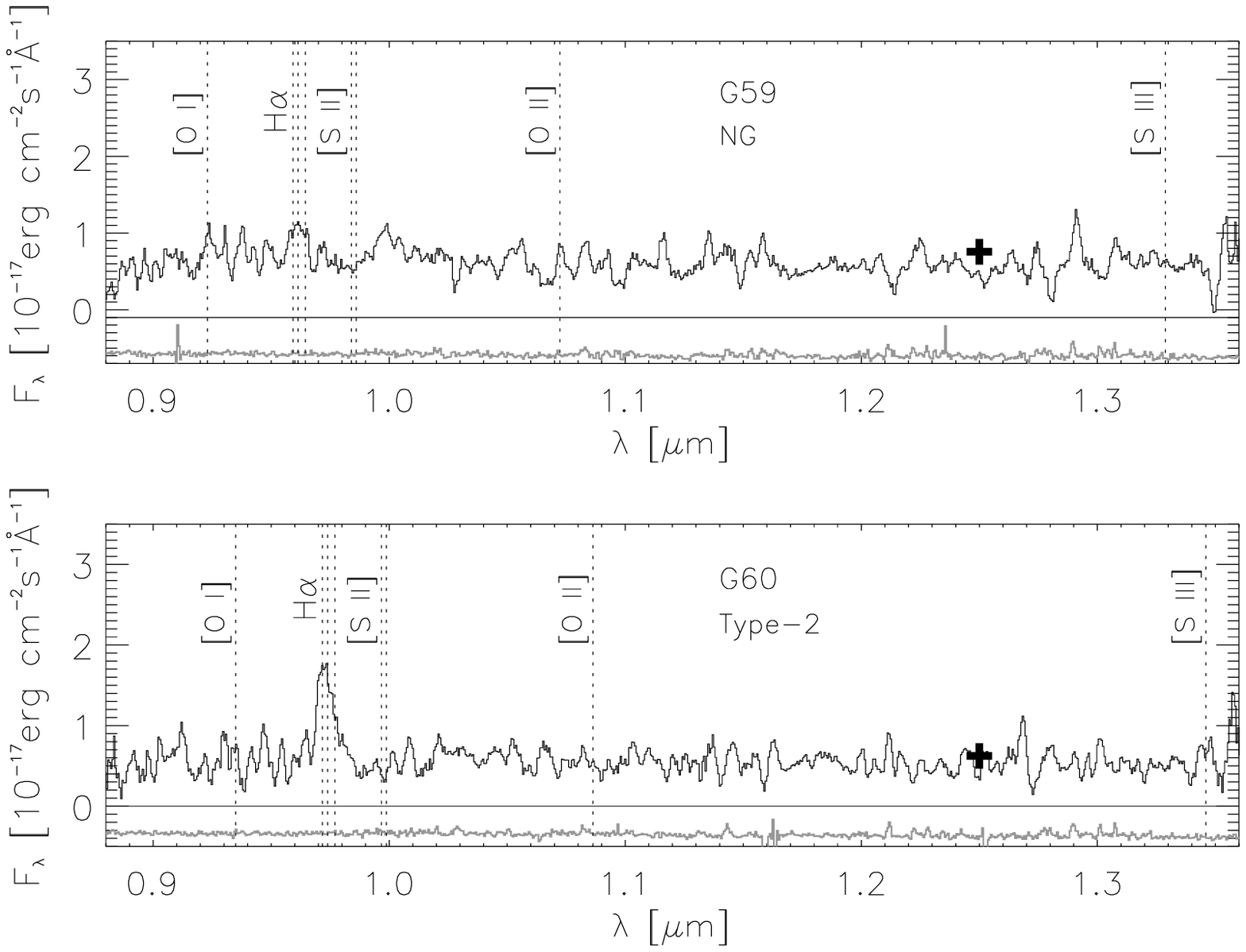}
\includegraphics[width=12cm]{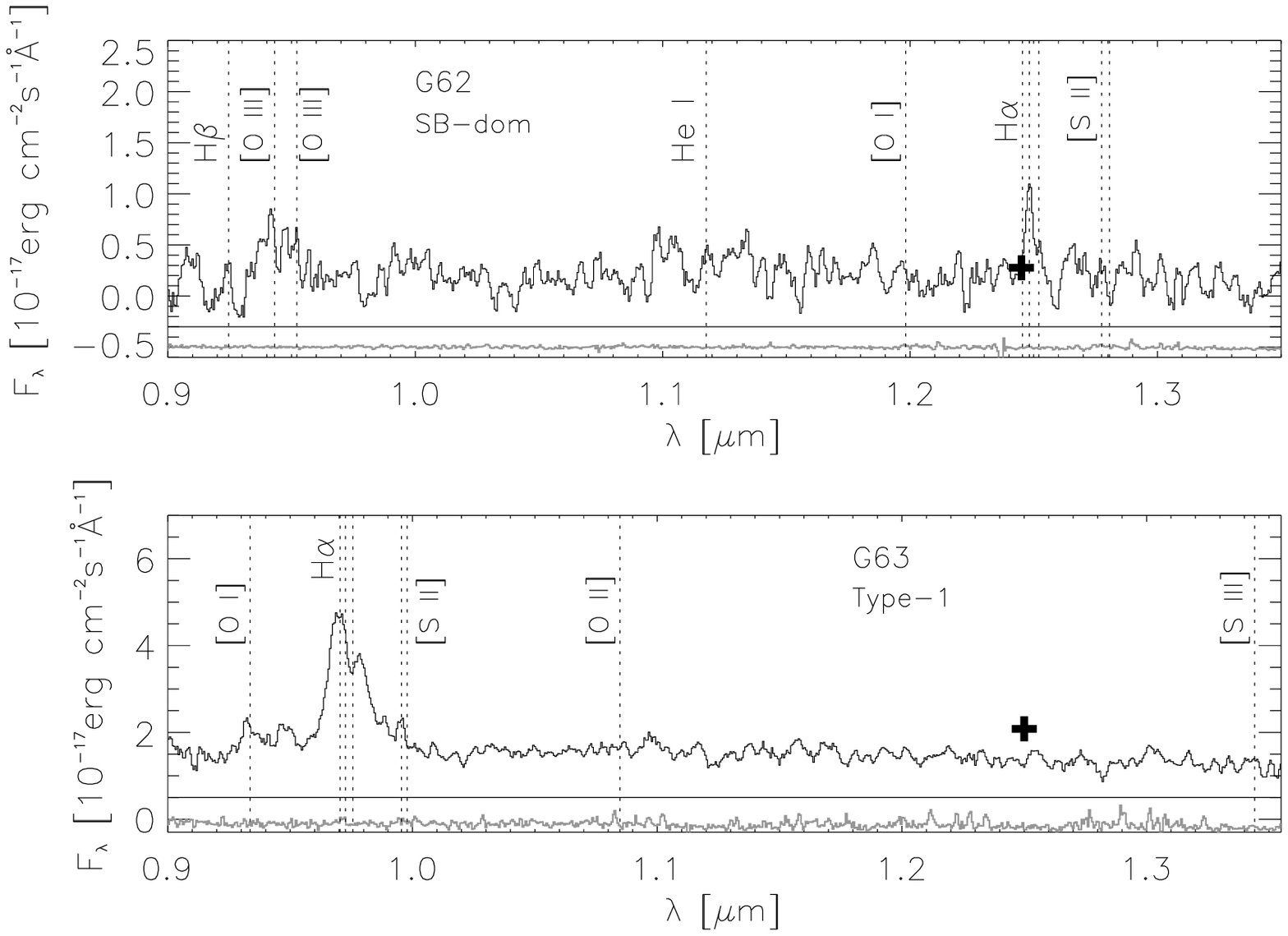}\par}
\caption{Same as in Figure \ref{groth5} but for the galaxies G59, G60, G62, and G63.}
\label{groth1a}
\end{figure*}

\begin{figure*}
\centering
{\par
\includegraphics[width=12cm]{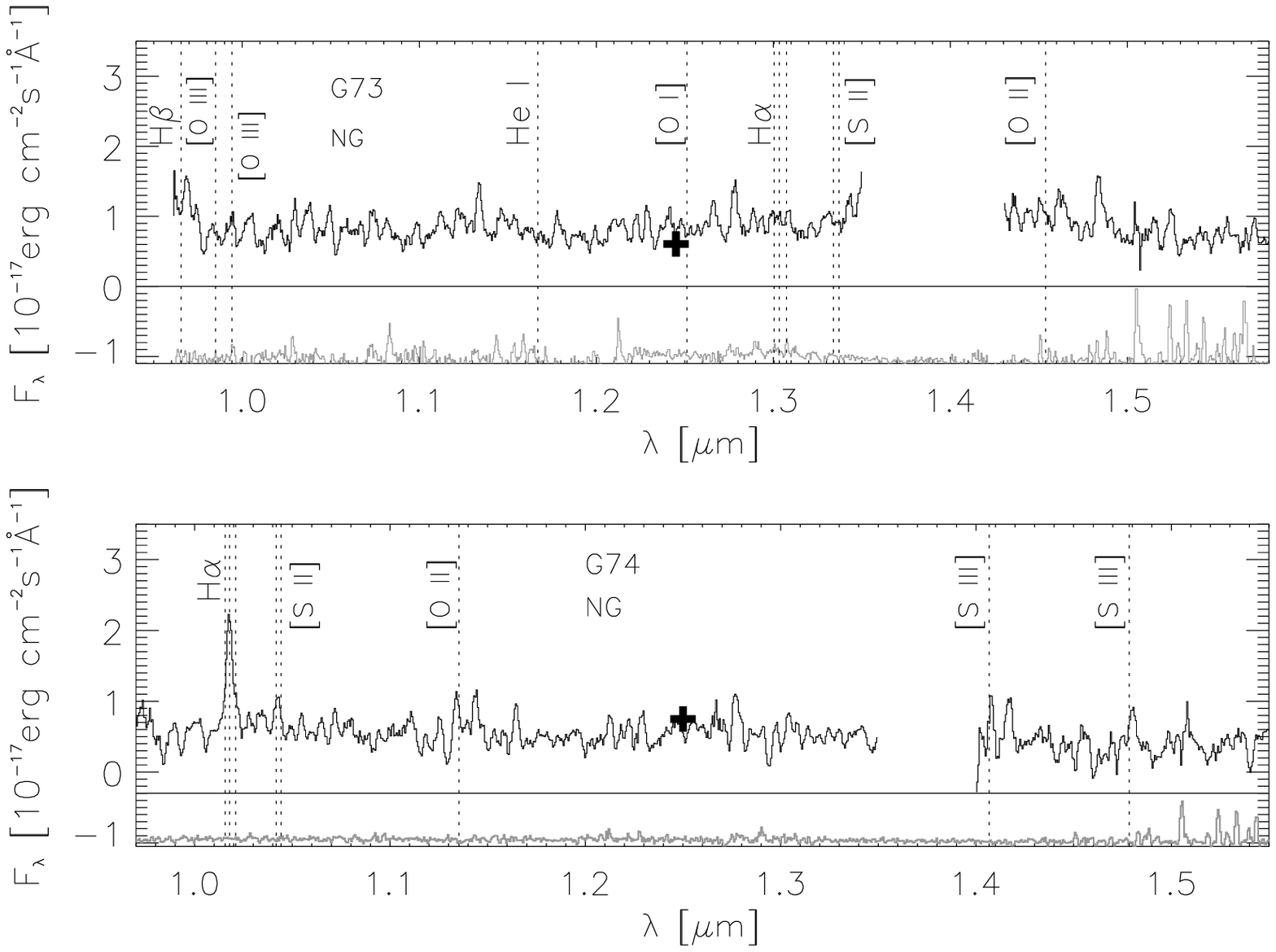}
\includegraphics[width=12cm]{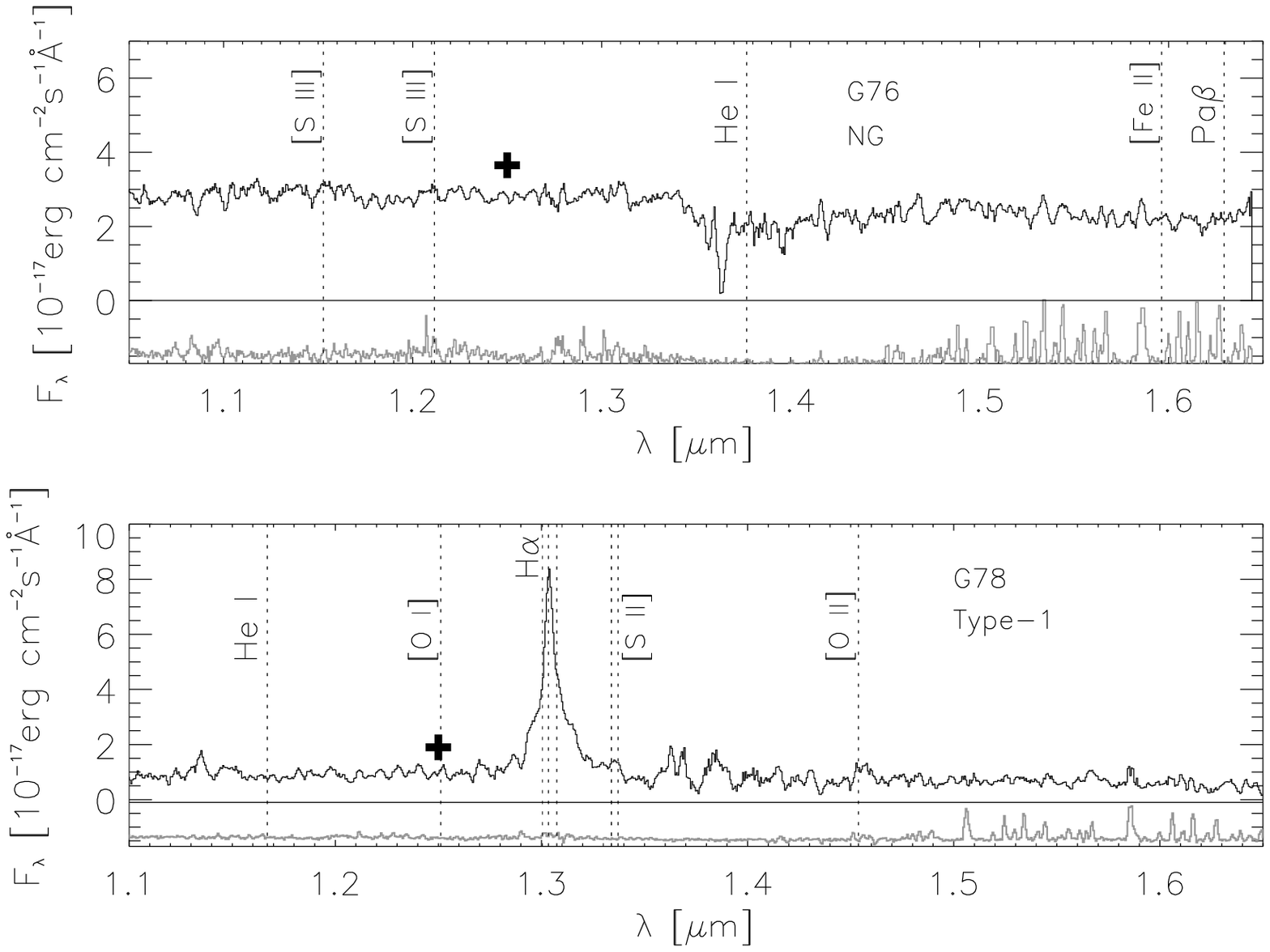}\par}
\caption{Same as in Figure \ref{groth5} but for the galaxies G73, G74, G76, and G78.}
\label{groth1b}
\end{figure*}

\begin{figure*}
\centering
{\par
\includegraphics[width=12cm]{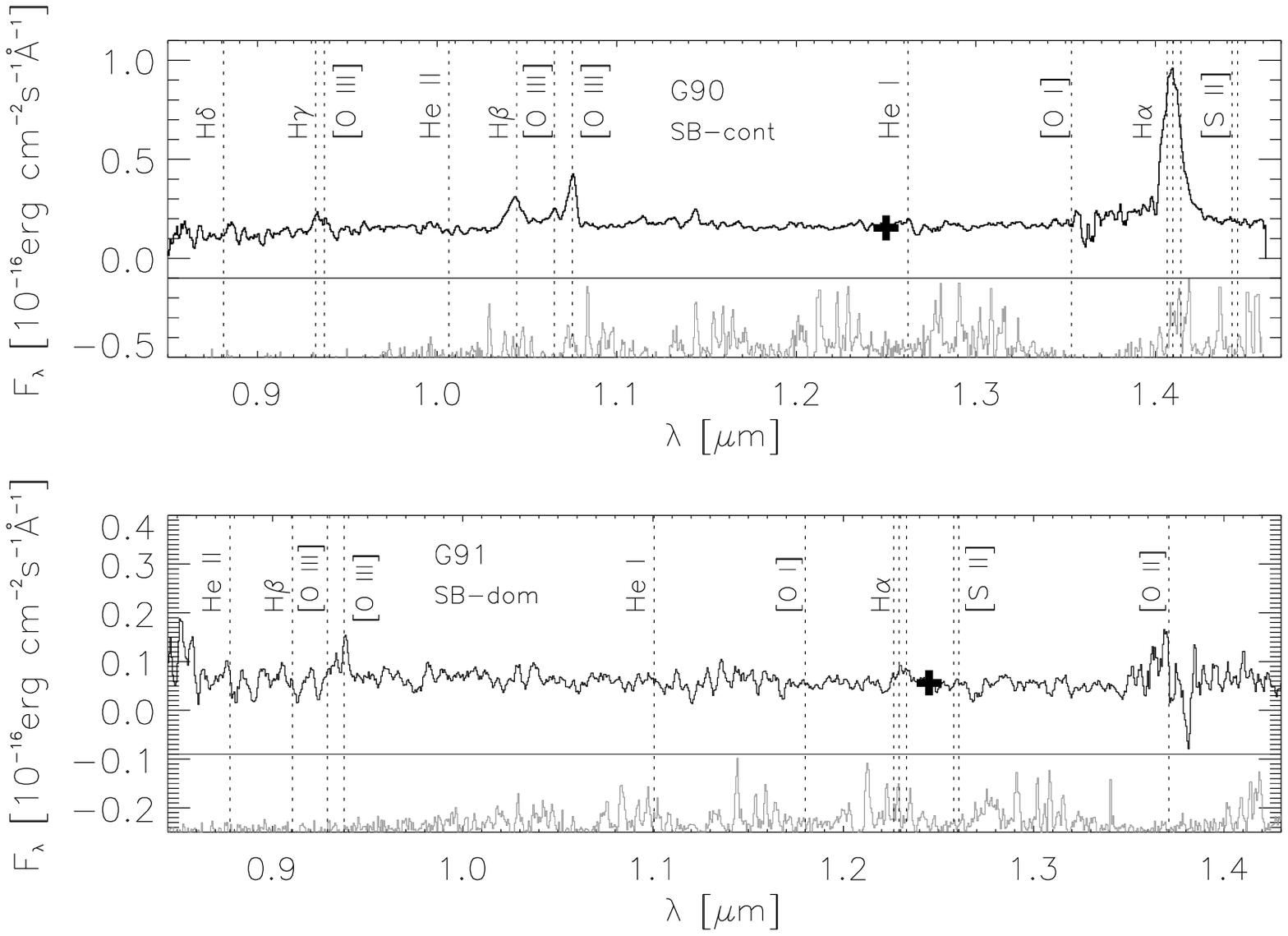}
\includegraphics[width=12cm]{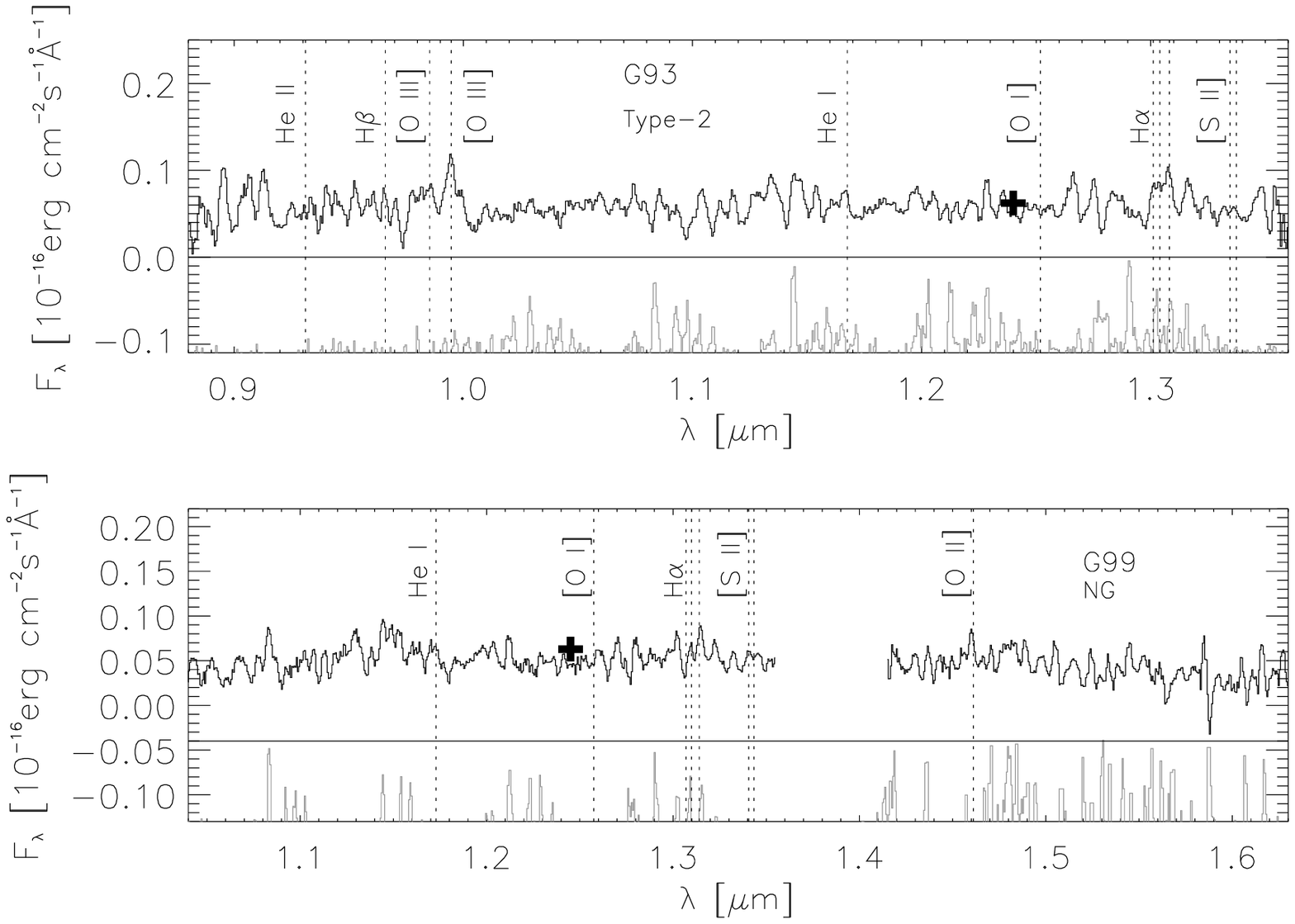}\par}
\caption{Same as in Figure \ref{groth5} but for the galaxies G90, G91, G93, and G99.}
\label{groth3a}
\end{figure*}

\begin{figure*}
\centering
{\par
\includegraphics[width=12cm]{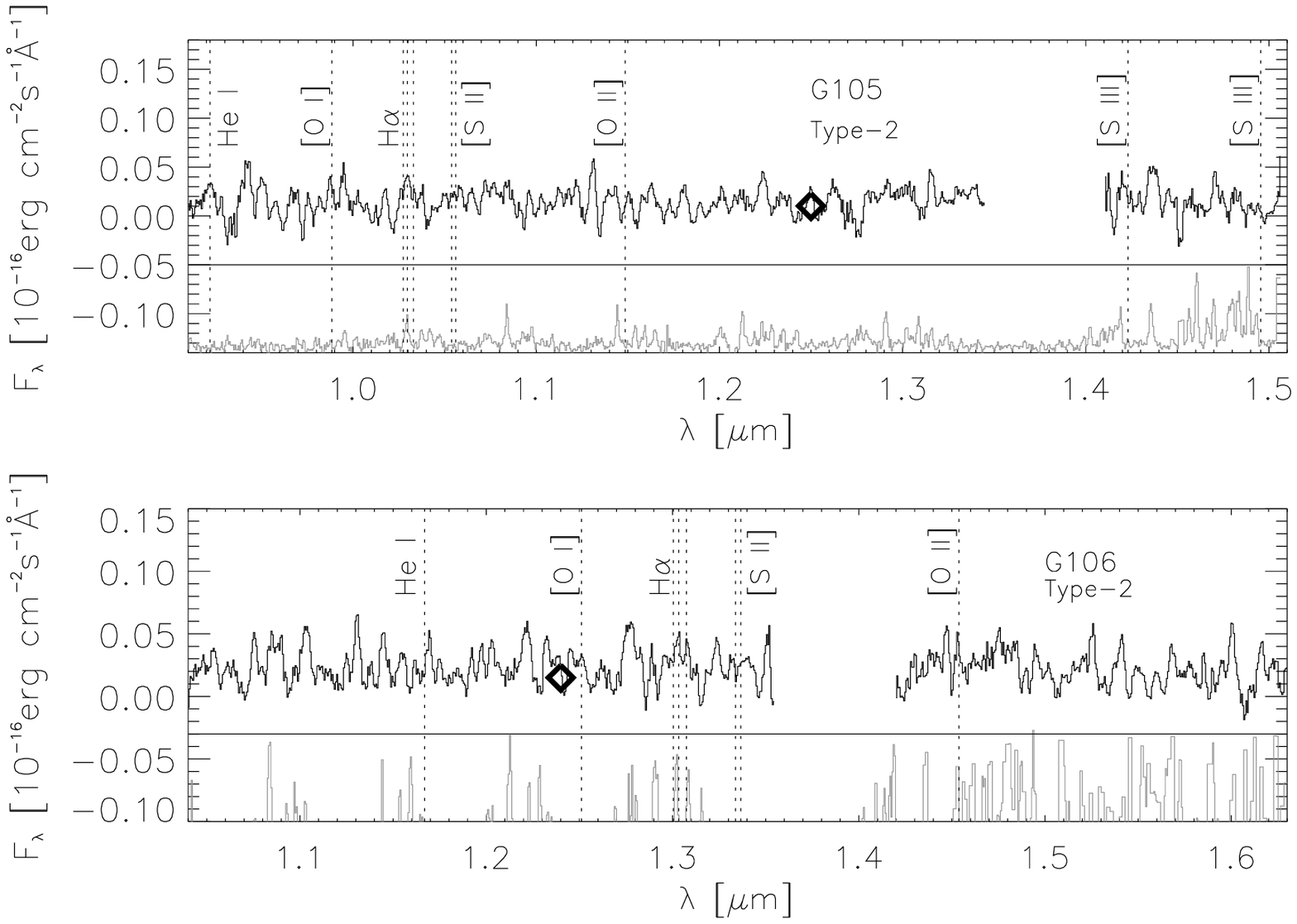}
\includegraphics[width=12cm]{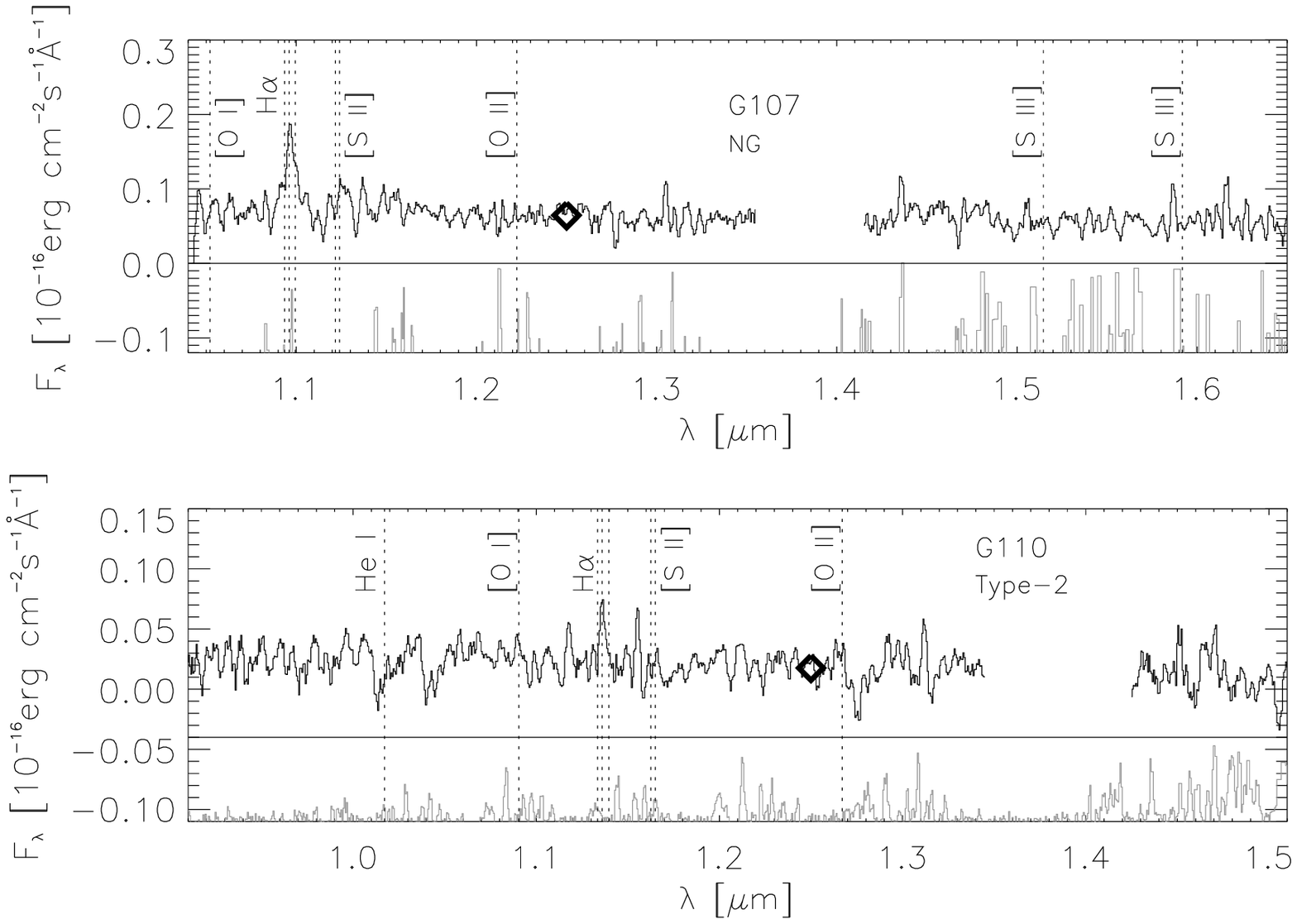}\par}
\caption{Same as in Figure \ref{groth5} but for the galaxies G105, G106, G107, and G110.}
\label{groth3b}
\end{figure*}

We estimated the uncertainties in the galaxy fluxes due to 
slit-losses for each mask using the reference stars, observed with the extra-slitlet 
mentioned above. We finally corrected the fluxes of the science targets by applying a correction factor 
to the individual spectra.  
The observed NIR spectra of the 28 AGN are shown in Figures \ref{groth5} to \ref{groth3b}. We have 
represented the observed J-band fluxes calculated from the Palomar-WIRC magnitudes reported in Table
\ref{photoz} for comparison with our flux-calibrated spectra. The Palomar fluxes were calculated in 
apertures of 2\arcsec~diameter \citep{Bundy06} and thus correspond to total fluxes for the majority 
of the galaxies in the sample (see Figure \ref{acs}). 
Consequently, these J-band fluxes should be equal or higher than the flux-calibrated LIRIS nuclear 
spectra, as indeed happens for all the galaxies. This gives us extra-confidence in our flux-calibration. 
For those galaxies without Palomar J-band magnitudes available (see Table \ref{photoz}), values from the 
SED fits shown in Figure \ref{seds} are given instead, and are represented as open diamonds. 
%The latter 
%values are also shown in Figures \ref{groth5} to \ref{groth3b} when the difference between the 
%observed and the SED J-band magnitudes are larger than 0.6 mag. 

The largest differences between the J-band photometry and spectroscopy are found for the galaxies G36, 
G43, and G47. The three of them are {\it NG} according to our SED fits, which are in agreement 
with their early-type morphologies, at least for G36 and G43 (those with ACS images available; see Figure \ref{acs}).
Thus, the nuclear flux of these objects (the spectra were extracted using apertures of 1\arcsec) will not 
be the dominant contribution to the total flux, as opposed to the AGN-dominated galaxies.
In addition, the three galaxies have relatively low redshifts: $z=[0.28,0.53]$, making more different
their nuclear and total emission. For example, 1\arcsec~at $z$=0.3 and 0.8 corresponds to 4.2 and 7.1 kpc respectively.

\subsection{DEEP optical spectroscopy}
\label{opt}

As indicated in Table \ref{photoz}, for 20 of the galaxies in the sample observed with LIRIS 
there are publicly available spectroscopic data from the Deep Extragalactic Evolutionary Probe 2 (DEEP2; \citealt{Davis03}).
The spectra were obtained during 80 nights between 2002 and 2005 using the DEIMOS multiobject spectrograph 
\citep{Faber03} on the Keck-II telescope, which 
provides a spectral resolution R = $\lambda$/$\Delta\lambda\sim5000$ covering the 
nominal range 6400-9100 \AA. DEEP2 employed a slit width of 1\arcsec, providing a 
%$\sim$68 km~s$^{-1}$ resolution at z=1.
spectral resolution of 1.4 \AA~and dispersion of 0.3 \AA~pixel$^{-1}$ \citep{Davis07}.
The reduced and extracted spectra were retrieved from the Keck website\footnote{http://tkserver.keck.hawaii.edu/egs/}
and they correspond to the Data Release 3 (DR3)\footnote{http://deep.berkeley.edu/DR3/}. 
Note that the Data Release 4 has recently become public \citep{Newman12}, but we have checked that 
there are no changes in the spectra of our galaxies. 
Thus, here we use the DR3 integrated spectra extracted along the locus of constant $\lambda$, 
following \citet{Horne86}, and corrected for the effects of tilted slits. 

The DEEP2 spectra are not flux calibrated and do not have the same spatial resolution as the LIRIS data. 
However, this can be solved by individually scaling the optical spectra to the averaged continuum flux of  
the overlapping (or bluest) NIR spectra. 
%, as shown in Figure \ref{scale} for the galaxy G60. 
The 20 flux-calibrated optical spectra are presented in Appendix \ref{appendix}. 
Our main interest in having optical spectra of our galaxies is to measure H$\beta$
fluxes for those galaxies with H$\alpha$ detected in the NIR, and thus determine 
individual values of the extinction to correct our H$\alpha$ fluxes. 

%\begin{figure*}
%\centering
%\includegraphics[width=12cm]{scale.ps}
%\caption{\footnotesize{Example of DEEP optical spectrum scaled to the LIRIS flux-calibrated spectrum of the galaxy G60.}
%\label{scale}}
%\end{figure*}

%Thus, we can measure the line fluxes and thus increase our wavelength 
%coverage. Line measurements are reported between brackets in Tables \ref{nir1}-\ref{otherlines}.  

\section{Results}

The NIR spectra of the sample observed with LIRIS show a wide variety of spectral features, including:
\begin{enumerate}
\item prominent emission lines typical of AGN and/or star forming galaxies (e.g. G17, G53, G55, G60, G78, and G90).
\item H$\alpha$ detected in absorption (G43 and G76).
\item Very weak emission lines or featureless spectra (e.g. G36, G99, and G106). 
\item Double-peaked emission lines corresponding to different kinematic components (G63).
\end{enumerate}

After scaling the DEEP2 spectra to the flux-calibrated LIRIS data, we measured the fluxes of the emission lines. 
We fitted Gaussian profiles using the 
Starlink program DIPSO. The resulting emission line fluxes and FWHMs of H$\alpha$, [N II]$\lambda$6584 \AA, H$\beta$, 
and [O III]$\lambda\lambda$4959,5007 \AA~are reported in Table \ref{nir1}. Same but for other detected 
transitions, such as [O I]$\lambda$6300 \AA, [O II]$\lambda$3727,7320 \AA, [O III]$\lambda$4363 \AA, and 
[S II]$\lambda$6732 \AA~is shown in Table \ref{otherlines}. 
The line flux errors reported in Tables \ref{nir1} and \ref{otherlines} are those given by DIPSO, 
and they include the uncertainty associated to the fit. 
All the line fluxes reported correspond to detections at $\geq$1.5$\sigma$ above the continuum noise, which 
we calculated as in Twite et al. (2012). We measure the 
continuum level and noise in two regions of $\sim$200 \AA~adjacent to a given 
emission line, and then calculate the 1.5$\times \sigma\times FWHM$ flux (hereafter 1.5$\sigma$) above the continuum. 
Emission line fluxes larger than that are quoted in Tables \ref{nir1} and \ref{otherlines}, 
otherwise we report an upper limit equal to the 1.5$\sigma$ flux. 
The exception are the sources lacking a secure spectroscopic redshift (G45, G57, G105, G106 and G110), 
for which a larger detection significance of 2$\sigma$ is required.

The majority of the emission lines are well 
reproduced with a single Gaussian of FWHM typical of the narrow-line region (NLR; $\la$ 1000 km~s$^{-1}$). 
%An example of this type of fit is shown in the top panel of Figure \ref{dipso}.
For seven of the galaxies, namely G17, G55, G60, G63, G78, G90, and G107, an additional
broad Gaussian component was needed to reproduce the observed line profiles of the permitted lines. 
%(see the middle panel of Figure \ref{dipso} for an example). 
The majority of these broad components have FWHM $\ga$ 2000 km~s$^{-1}$, typical of lines originated 
in the broad-line region (BLR). However, for some of the galaxies, intermediate-width components 
have been fitted instead (FWHM $\sim$ 1000--1500 km~s$^{-1}$).

Finally, in some cases it was necessary to fit two Gaussians to reproduce the profiles of the narrow lines 
in the DEEP2 spectra, of higher spectral resolution than the LIRIS data. 
%(see bottom panel of Figure \ref{dipso}). 
This is the case of the galaxies G53, G55, G62, G63, G74, G93 and G99. In the LIRIS spectrum of G63, the double 
component of the H$\alpha$ line is clearly resolved as well (see Figure \ref{groth1a}).

%\begin{figure*}
%\centering
%{\par
%\includegraphics[width=7cm,angle=-90]{f11a.eps}
%\includegraphics[width=7cm,angle=-90]{f11b.eps}
%\includegraphics[width=7cm,angle=-90]{f11c.eps}\par}
%\caption{\footnotesize{Examples of line fits using narrow components only (G53; H$\alpha$+[N II] and 
%[S II]$\lambda\lambda$6717, 6732; top), narrow and broad components (G78; H$\alpha$+[N II]; center), 
%and two narrow components (G74, [O III]$\lambda\lambda$4959, 5007; bottom). The X- and Y-axis are given
%in the same units as in Figure \ref{groth5}, except for the bottom panel, for which the X-axis in given
%in angstroms.}
%\label{dipso}}
%\end{figure*}

%\clearpage

\begin{table*}
\centering
\scriptsize
\begin{tabular}{lcclcccccccccc}
\hline
\hline
ID &\multicolumn{2}{c}{z$_{spec}$} & Comp. &\multicolumn{2}{c}{H$\alpha$} &\multicolumn{2}{c}{[N II]$\lambda$6584}&\multicolumn{2}{c}{H$\beta$}  &\multicolumn{2}{c}{[O III]$\lambda$4959}&\multicolumn{2}{c}{[O III]$\lambda$5007}\\
            &z$_{opt}$ &z$_{NIR}$&       &Flux& FWHM & Flux& FWHM&Flux& FWHM&Flux& FWHM &Flux& FWHM  \\
\hline
17  &  1.284  & 1.284 & N &   \dots         & \dots    &  \dots	         & \dots     & 1.06$\pm$1.52   & 820	   &  10.1$\pm$1.9    &  810	&  21.2$\pm$2.0  & 800      \\   
    &         &       & B &   \dots         & \dots    &  \dots	         & \dots     & 55.7$\pm$4.93   & 3400	   &  ...	      &  ...	&   ... 	 &\dots     \\   
25  &  0.761  & 0.761 & N &  1.48$\pm$0.24  &$\leq$340 &  1.16$\pm$0.21  &$\leq$340  & [0.25$\pm$0.02] & [140]     & [:0.07]          & [130]	& [0.17$\pm$0.02]& [130]    \\ 
26  &  0.808  & 0.808 & N &  1.03$\pm$0.33  &$\leq$280 &  0.72$\pm$0.20  &$\leq$250  & [0.09$\pm$0.02] & [70]      & [:0.05]          & [70]	& [0.10$\pm$0.01]&  [70]    \\ 
27  &  0.683  & -     & N &  :0.60          &  \dots   &  :0.60  	 & \dots     & [0.18$\pm$0.04] & [50]	   & [0.29$\pm$0.02]  &[$\leq$80]&[0.89$\pm$0.05]&[$\leq$80]\\
\hline
36  &  0.281  & -     & N &  [0.66$\pm$0.02]& [280]    & [0.34$\pm$0.01] &  [280]    &  \dots	       &  \dots    &  \dots	      &  \dots  & \dots 	 &\dots     \\         
43  &  0.532  & 0.534 & N &  Abs            &  720     & 0.6$\pm$0.3	 &  720      & [Abs]	       & [700]     &  [:0.08]	      &  \dots  &  [:0.08]	 &\dots     \\   
45  &    -    & 1.268 & N &  :0.33          &$\leq$190 & 0.48$\pm$0.13   & $\leq$190 & Abs	       & $\leq$180 & :0.27            &$\leq$250&  0.55$\pm$0.13 &$\leq$250 \\  
47  &  0.418  & 0.418 & N &  \dots          &  \dots   &   \dots	 & \dots     &  \dots	       &  \dots    &  \dots	      &  \dots  &\dots  	 &\dots     \\
53  &  0.723  & 0.722 & Nb&  1.47$\pm$0.11  & $\leq$340& 0.22$\pm$0.09   &$\leq$340  &[0.08$\pm$0.01]  &[$\leq$75] &[:0.06]           &[$\leq$70]&[:0.06]        &[$\leq$70]\\   
    &         &       & Nr&  \dots          & \dots    &   \dots	 &  \dots    &[0.13$\pm$0.02]  &[$\leq$75] &[:0.06]           &[$\leq$70]&[0.10$\pm$0.01]&[$\leq$70]\\   
55  &  1.211  & 1.212 & N &  0.5$\pm$0.2    & $\leq$180& 0.27$\pm$0.20   & $\leq$180 &     :0.23       & \dots     &  :0.23	      &  \dots  &	 :0.23   &\dots     \\        
    &         &       & B &  :2.02          &  1460    &  \dots 	 &\dots      &   :1.65         & \dots     &  \dots	      &  \dots  & \dots 	 &\dots     \\
56  &  1.208  & 1.213 & N &  0.88$\pm$0.24  &  180     &  :0.33          & 180       &   :0.20         & \dots     &   :0.20	      &  \dots  & :0.20 	 &\dots     \\  
57  &    -    & 0.651 & N &  \dots          &  \dots   &   \dots	 &\dots      &  \dots	       & \dots     &   \dots	      &  \dots  &\dots  	 &\dots     \\
\hline
59  &  0.465  & 0.465 & N &  1.42$\pm$0.50  & 230      &  1.05$\pm$0.47  & 230       & [0.32$\pm$0.02] & [220]     & [1.09$\pm$0.03]  & [210]	& [3.46$\pm$0.04]&[210]     \\
60  &  0.484  & 0.484 & N &  1.14$\pm$1.10  & 230      &   1.6$\pm$0.6   & 230       & [0.34$\pm$0.02] & [170]     & [0.30$\pm$0.02]  & [160]	& [1.02$\pm$0.03]&[160]     \\ 
    &         &       & B &  6.09$\pm$1.66  & 1510     &  \dots 	 &\dots      & [0.91$\pm$0.08] & [1530]    &   \dots	      &  \dots  & \dots 	 &\dots     \\ 
62  &  0.902  & 0.902 & N &  2.87$\pm$0.46  & 50       & 1.1$\pm$0.4	 & 50	     &  :0.97          & 40	   &  :0.97           &  40	&1.67$\pm$0.46   & 40	    \\ 
63  &  0.482  & 0.482 & Nb&  4.82$\pm$0.98  & 590      & 4.67$\pm$1.25   & 590       & [0.07$\pm$0.02] & [880]     & [0.23$\pm$0.03]  & [850]	& [0.69$\pm$0.11]& [840]    \\ 
    &         &       & Nr&  2.39$\pm$0.68  & 230      & 1.33$\pm$0.64   & 230       & [0.10$\pm$0.03] & [180]     & [0.52$\pm$0.03]  & [310]	& [1.57$\pm$0.08]& [310]    \\ 
    &         &       & B &  35.7$\pm$4.0   & 5530     & \dots  	 &\dots      & [12.7$\pm$1.5]  & [5880]    &  \dots	      &  \dots  & \dots 	 &\dots     \\ 
73  &  0.986  & -     & N &  :0.79          &\dots     & :0.79		 &\dots      & :1.13	       & \dots     &  :1.13	      &  \dots  &  :1.13	 &\dots     \\
74  &  0.551  & 0.551 & Nb&  5.25$\pm$0.53  & 280      &1.29$\pm$0.44	 & 280       &[0.53$\pm$0.07]  & [120]     &[0.40$\pm$0.03]   & [65]	&[1.2$\pm$0.1]   & [65]     \\
    &         &       & Nr&   \dots         &\dots     &\dots		 &\dots      &[0.49$\pm$0.07]  & [120]     &[0.36$\pm$0.03]   & [150]	&[1.09$\pm$0.08] & [150]    \\
76  &  0.271  & -     & N &  [Abs]          & [210]    &[0.9$\pm$0.2]	 & [210]     & \dots	       & \dots     &  \dots	      &  \dots  &    \dots	 &\dots     \\
78  &  0.985  & 0.986 & N &  17.97$\pm$0.13 & 360      & 3.15$\pm$1.56   & 410       & \dots	       & \dots     &  \dots	      &  \dots  &   \dots	 &\dots     \\
    &         &       & B &  62.91$\pm$0.34 & 3180     &  \dots 	 &\dots      & \dots	       & \dots     &  \dots	      &  \dots  &	\dots	 &\dots     \\
\hline
90  &  1.148  & 1.148 & N & 19.4$\pm$3.2    & 640      & 16.4$\pm$4.7	 & 640       & 2.58$\pm$1.33   & 780	   &4.78$\pm$0.67     &  760	& 14.3$\pm$0.8   & 750      \\
    &         &       & B & 52.6$\pm$8.9    & 1830     &  \dots 	 &  \dots    &  12$\pm$2       & 1820	   &  \dots	      &   \dots & \dots 	 &\dots     \\
91  &  0.873  & 0.873 & N & 1.07$\pm$0.27   & $\leq$260& 1.07$\pm$0.25   & $\leq$260 &   Abs	       & 360	   &   1.47$\pm$0.38  & 350	&2.61$\pm$0.48   & $\leq$310\\
93  &  0.985  & 0.987 & N & 2.44$\pm$0.83   & 130      & 3.58$\pm$0.79   & 130       &  :0.81          & 120	   &   :0.81          & 120	&2.05$\pm$0.47   & 120      \\
99  &  0.996  & 0.996 & N & :0.61           & \dots    &  1.1$\pm$0.3  	 & \dots     &  \dots	       &\dots	   &  \dots	      &   \dots &\dots  	 &\dots     \\
105 &    -    & 0.569 & N & 1.01$\pm$0.35   & $\leq$370& :0.68           & $\leq$370 &  \dots	       &\dots	   &  \dots	      &   \dots &\dots  	 &\dots     \\
106 &    -    & 0.986 & N & :1.03           & 95       & :1.03           & 95	     &  \dots	       &\dots	   &  \dots	      &   \dots &\dots  	 &\dots     \\ 
107 &  0.671  & 0.670 & N &  2.6$\pm$0.9    & 90       & 1.22$\pm$0.57   & 90	     &  [0.37$\pm$0.07]& [$\leq$90]&[:0.37]           &[$\leq$85]&[:0.37]        &[$\leq$85]\\
    &         &       & B & 2.05$\pm$1.21   & 870      &  \dots 	 &  \dots    &  [:3.12]        & [850]     &  \dots	      &   \dots &\dots  	 &\dots     \\
110 &    -    & 0.731 & N & 1.74$\pm$0.47   & 270      & :0.69           & 270       &  \dots	       &\dots	   &  \dots	      &   \dots &\dots  	 &\dots     \\
\hline
\end{tabular}
\caption{Line fluxes and FWHMs. Columns 2 and 3 list the spectroscopic redshifts from optical and NIR spectra, respectively. 
Column 4 corresponds to the line component that has been fitted (B: broad, N: narrow, Nb: blueshifted narrow, and Nr: redshifted narrow component).
Lines fluxes ($\times$10$^{-16}~erg~s^{-1}~cm^{-2}$) and FWHMs (km~s$^{-1}$) of H$\alpha$, [N II]$\lambda$6584, H$\beta$, and [O III]$\lambda\lambda$4959, 5007 
are reported. The line flux errors reported are given by DIPSO, and they include the uncertainty associated to the fit.
For several faint/undetected transitions, the continuum noise was used to calculate a 1.5$\sigma$ upper limit to the line emission (those starting with :). 
For the sources lacking a secure spectroscopic redshift (G45, G57, G105, G106 and G110), a larger detection significance of 2$\sigma$ is required. 
%%When line fluxes have been measured, but line flux is below the 1.5$\sigma$ upper limit, the FWHM of the line profile is given. 
Brackets indicate that measurements are from DEEP2 optical spectra.
Line widths have been corrected from instrumental broadening (18 and 3.6 \AA~for the LIRIS and DEEP2 spectra respectively). Typical line width uncertainties 
are 25--30\%.}
\label{nir1}
\end{table*}

\begin{table}
%\centering 
\begin{tabular}{lcccc}
%%\tablecaption{Other line fluxes and FWHMs\label{otherlines}}
\hline
\hline
ID & Comp. & Line & Flux & FWHM \\
\hline
36  & N & [S II]$\lambda$6732  & [0.17$\pm$0.01] & [260]              \\
43  & N & [O II]$\lambda$7320  & 0.31$\pm$0.17   & 530                \\
45  & N & H$\gamma$            & Abs             & 150                \\
    & N & [O III]$\lambda$4363 & 0.34$\pm$0.10   & $\leq$280          \\
47  & N & [O II]$\lambda$7320  & 0.46$\pm$0.14   & 220                \\
53  & N & [S II]$\lambda$6732  & 0.31$\pm$0.30   & $\leq$340          \\
55  & Nb& [O II]$\lambda$3727  & [0.09$\pm$0.01] & [95]               \\
    & Nr&                      & [0.10$\pm$0.02] & [95]               \\
56  & N & [O II]$\lambda$3727  & [0.17$\pm$0.03] & [170]              \\  
57  & N & [O II]$\lambda$7320  & 2.79$\pm$0.24   & 180                \\
\hline 
59  & N & [O II]$\lambda$7320  & 1.28$\pm$0.40   & 210                \\
62  & Nb& [O II]$\lambda$3727  & [0.24$\pm$0.11] & [130]              \\ 
    & Nr&                      & [0.50$\pm$0.12] & [160]              \\ 
63  & N & [O I]$\lambda$6300   & 2.09$\pm$0.97   & 470                \\ 
74  & N & [S II]$\lambda$6732  & 0.86$\pm$0.32   & 270                \\
\hline
90  & N & [O II]$\lambda$3727  & [1.63$\pm$0.10] & [490]              \\      
91  & N & [O II]$\lambda$3727  & [0.59$\pm$0.05] & [320]              \\ 	   
93  & Nb& [O II]$\lambda$3727  & [0.14$\pm$0.20] & [240]              \\ 	   
    & Nr&                      & [0.51$\pm$0.20] & [320]              \\ 
99  & N & [O II]$\lambda$7320  &  1.24$\pm$0.54  & 170                \\	  
    & N & [S II]$\lambda$6732  &  :0.61          & 170                \\   
    & Nb& [O II]$\lambda$3727  &  [0.2$\pm$0.03] & [120]              \\
    & Nr&                      &  [0.2$\pm$0.04] & [120]              \\                
110 & N &  [O I]$\lambda$6300  &  0.72$\pm$0.41  & 280                \\ 		   
\hline
\end{tabular}
\caption{Same as in Table \ref{nir1}, but for other lines detected in the LIRIS 
and/or DEEP2 spectra.}
\label{otherlines}
\end{table}

In Figures \ref{groth5} to \ref{groth3b} we show a sky spectrum for each of the galaxies, extracted 
using the same aperture as for the nuclear spectra. Although the sky lines are removed from the spectra
of the targets, 
as we described in Section \ref{nir}, residual noise could be affecting the emission line fluxes reported in Tables
\ref{nir1} and \ref{otherlines}. In particular, the galaxies whose H$\alpha$ emission can be somehow affected 
by contamination of sky lines are G45, G55, G56, G90, G91, G93, G99, and G106.

\subsection{Comparison between SED-fitting and spectral classification}
\label{comparison}

On the basis of the spectroscopic features detected for the 28 galaxies observed with LIRIS, we can 
classify them spectroscopically to compare with the 
SED classification and to confirm the presence of nuclear activity.
In Table \ref{comparison_broad} we summarize the results from this comparison.

\begin{table*}
\centering
\begin{tabular}{lcccclcc}
\hline
\hline
%\tablecaption{Comparison between spectroscopic data and SED fitting. \label{comparison_broad}}
ID & z$_{spec}$ & Comp. & \multicolumn{2}{c}{H$\alpha$ broad} & Group &  Agreement & AGN \\
    &        &            & FWHM (km~s$^{-1}$) & Log L (erg~s$^{-1}$) & & & \\
\hline
17  & 1.284  &  N, B      & 3400 (H$\beta$) & 43.69 (H$\beta$)& Type-1  & $\surd$ &$\surd$	\\
25  & 0.761  &  N         & \dots  & \dots  & NG           & $\surd$ &$\surd$	 \\
26  & 0.808  &  N         & \dots  & \dots  & SB-dom       & $\surd$ &$\surd$	 \\
27  & 0.683  &  N         & \dots  & \dots  & SB-cont      & $\surd$ &$\surd$	 \\
\hline
36  & 0.281  &  N         & \dots  & \dots  & NG           & $\surd$ &$\surd$	 \\
43  & 0.532  &  N         & \dots  & \dots  & NG           & $\surd$ &$\surd$	 \\
45  & 1.268  &  N         & \dots  & \dots  & SB-dom       & $\surd$ &$\surd$	 \\
47  & 0.418  &  N         & \dots  & \dots  & NG           & $\surd$ & ?	 \\
53  & 0.723  &  Nb, Nr    & \dots  & \dots  & Type-2       & X       &$\surd$	 \\
55  & 1.212  &  Nb, Nr, B & 1500   &$\leq$42.19  & SB-dom  & $\surd$ & ?	 \\
56  & 1.208  &  N         & \dots  & \dots  & Type-2       & $\surd$ &$\surd$	 \\  
57  & 0.651  &  N         & \dots  & \dots  & Type-2       & $\surd$ &$\surd$	 \\
\hline
59  & 0.465  &  N         & \dots  & \dots  & NG           & X       &$\surd$	 \\
60  & 0.484  &  N, B   	  & 1500   & 41.68  & Type-2       & X       &$\surd$	 \\ 
62  & 0.902  &  Nb, Nr    & \dots  & \dots  & SB-dom       & $\surd$ & ?	 \\ 
63  & 0.482  &  Nb, Nr, B & 5500   & 42.45  & Type-1       & $\surd$ &$\surd$	 \\ 
73  & 0.986  &  N         & \dots  & \dots  & NG           & $\surd$ &$\surd$	 \\
74  & 0.551  &  Nb, Nr    & \dots  & \dots  & NG           & $\surd$ &$\surd$	 \\
76  & 0.271  &  N         & \dots  & \dots  & NG           & $\surd$ &$\surd$	 \\
78  & 0.986  &  N, B      & 3200   & 43.46  & Type-1       & $\surd$ &$\surd$	 \\
\hline
90  & 1.148  &  N, B      & 1800   & 43.55  & SB-cont      & $\surd$ &$\surd$	 \\
91  & 0.873  &  N      	  & \dots  & \dots  & SB-dom       & $\surd$ &$\surd$	 \\
93  & 0.987  &  N      	  & \dots  & \dots  & Type-2       & $\surd$ &$\surd$	 \\
99  & 0.996  &  Nb, Nr    & \dots  & \dots  & NG           & $\surd$ &$\surd$	 \\
105 & 0.569  &  N      	  & \dots  & \dots  & Type-2       & $\surd$ &$\surd$	 \\
106 & 0.986  &  N      	  & \dots  & \dots  & Type-2       & $\surd$ &$\surd$	 \\
107 & 0.670  &  N, B      & 900    & 41.55  & NG           & $\surd$ & ?	 \\
110 & 0.731  &  N      	  & \dots  & \dots  & Type-2       & $\surd$ &$\surd$	 \\
\hline
\end{tabular}
\caption{Comparison between spectroscopic data and SED fitting. Column 2 corresponds to z$_{spec}$ measured from the NIR and optical lines 
(see Table \ref{nir1}). In case of disagreement, we chose the 
most reliable value in terms of the S/N of the lines identified in each case. Columns 3, 4, and 5 indicate the emission line components fitted 
(B: broad, N: narrow, Nb: blueshifted narrow, and 
Nr: redshifted narrow component) and the FWHM and luminosity of the broad component of H$\alpha$ (H$\beta$ in the case of G17), if present. 
Column 6 lists the SED-based classification of the galaxies as in \citealt{Ramos09}. Finally, 
columns 7 and 8 indicate whether or not there is agreement between the spectral and SED classifications, and if the presence of nuclear activity 
has been confirmed from the different diagnostic diagrams employed here.}
\label{comparison_broad}
\end{table*}

First, we have to confirm the presence of nuclear activity in all the sources.
At the flux limits of the X-ray surveys from which the sources were selected, most of the 
galaxies are expected to be AGN and have Log (f$_X/f_{opt})>$ -1, indicating that 
they are not quiescent (i.e. non-active) galaxies. Thus, in \citealt{Ramos09} we assumed that 
all the X-ray/MIR sources were AGN. However, it is also possible that some of them are simply
star-forming galaxies emitting in the X-ray. Thus, considering that here we have spectroscopic information 
for 28 galaxies, in addition to the SED classification and X-ray luminosities, 
it is worth confirming that they are AGN. 

The first thing to check are the hard X-ray luminosities. According to \citet{Ranalli12}, 
the criterium for a secure AGN origin of the X-ray luminosity is L$_X > 10^{42}(1+z)^{2.7}$ erg~s$^{-1}$ = 
L$_X^{AGN}$. However, this criterium is extremely restrictive, since it ignores the existence of low-luminosity AGN.
As an example, in \citet{Pereira11} the authors studied the X-ray emission of a sample of 27 local luminous infrared 
galaxies (LIRGs) and found that the largest 2-10 keV luminosity was 2$\times10^{41}$ erg~s$^{-1}$. Such L$_X$ corresponds to 
a LIRG that is forming stars at a rate of 92 M$_{\sun}~yr^{-1}$.  
Only 8/28 sources have L$_X<L_X^{AGN}$, namely G43, G47, G55, G59, G62, G76, G105 and G107. Thus, for these galaxies, 
either their X-ray and MIR emission are due to intense star formation, or to a heavily obscured AGN.

Secondly, we will contrast the previous information with the spectroscopic and 
SED classifications. The three galaxies fitted with {\it Type-1} templates 
(G17, G63, and G78; see Figure \ref{seds}) show broad components 
in their optical and/or NIR spectra. In fact, these are the only three galaxies with broad lines of FWHM $>$ 3000 km~s$^{-1}$.  
G90 shows broad H$\alpha$ and H$\beta$ components of FWHM $\sim$ 1800 km~s$^{-1}$, in agreement with its 
SED classification as a {\it SB-cont}: the fitted template for this galaxy is
a Sy1/starburst/ULIRG SED. On the other hand, there are another three galaxies with broad lines detected in their spectra 
(G55, G60, and G107) which were not classified as Type-1 AGN from their SED fits. Note, however, that those broad
components have FWHMs $\la$1500 km~s$^{-1}$, and the luminosities of their broad H$\alpha$ components are considerably
lower than those of the galaxies G63, G78, and G90, as we show in Table \ref{comparison_broad}. In fact, 
broad H$\alpha$ emission is also detected in starburst galaxies (up to 2200 km~s$^{-1}$ in the case of 
stellar winds or supernovae remnants; see e.g. \citealt{Boyle95,Homeier99,Westmoquette11}). Thus, G55 and G107 
could be either normal galaxies with intense star formation, or heavily obscured Type-1 AGN.

For G43 and G76 we detect H$\alpha$ in absorption (as well as H$\beta$ in the case of G43), 
which is characteristic of early-type galaxies. This agrees with our SED classification as {\it NG} 
(see Table \ref{comparison_broad}) and shows how deeply buried --or possibly absent-- 
these AGN are. Of the other 
galaxies classified in this group, some of them show narrow permitted lines
(G25, G36, G59, and G74) and the rest have either very faint emission lines or practically featureless spectra 
(G47, G73, and G99).

The galaxies classified as {\it SB-dom} (G26, G45, G62, and G91) show very similar spectra, with relatively 
faint H$\alpha$ emission and H$\beta$ either in absorption or detected with very low signal-to-noise. Finally, the 
galaxies fitted with {\it Type-2} templates show narrow emission lines (G53, G56, G57, G93, G105, G106, and G110), as
well as the {\it SB-cont} G27, whose SED was fitted with a Seyfert 2/starburst template in \citealt{Ramos09}. 
%Summarizing, the classification method presented in \citealt{Ramos09} based on optical-to-MIR SED fitting successfully
%identifies Type-1 AGNs with strong broad-line components but fails identifying Type-1 AGNs with broad components
%of FWHMs $\la$1500 km~s$^{-1}$.

\begin{figure*}
\centering
\includegraphics[width=10cm]{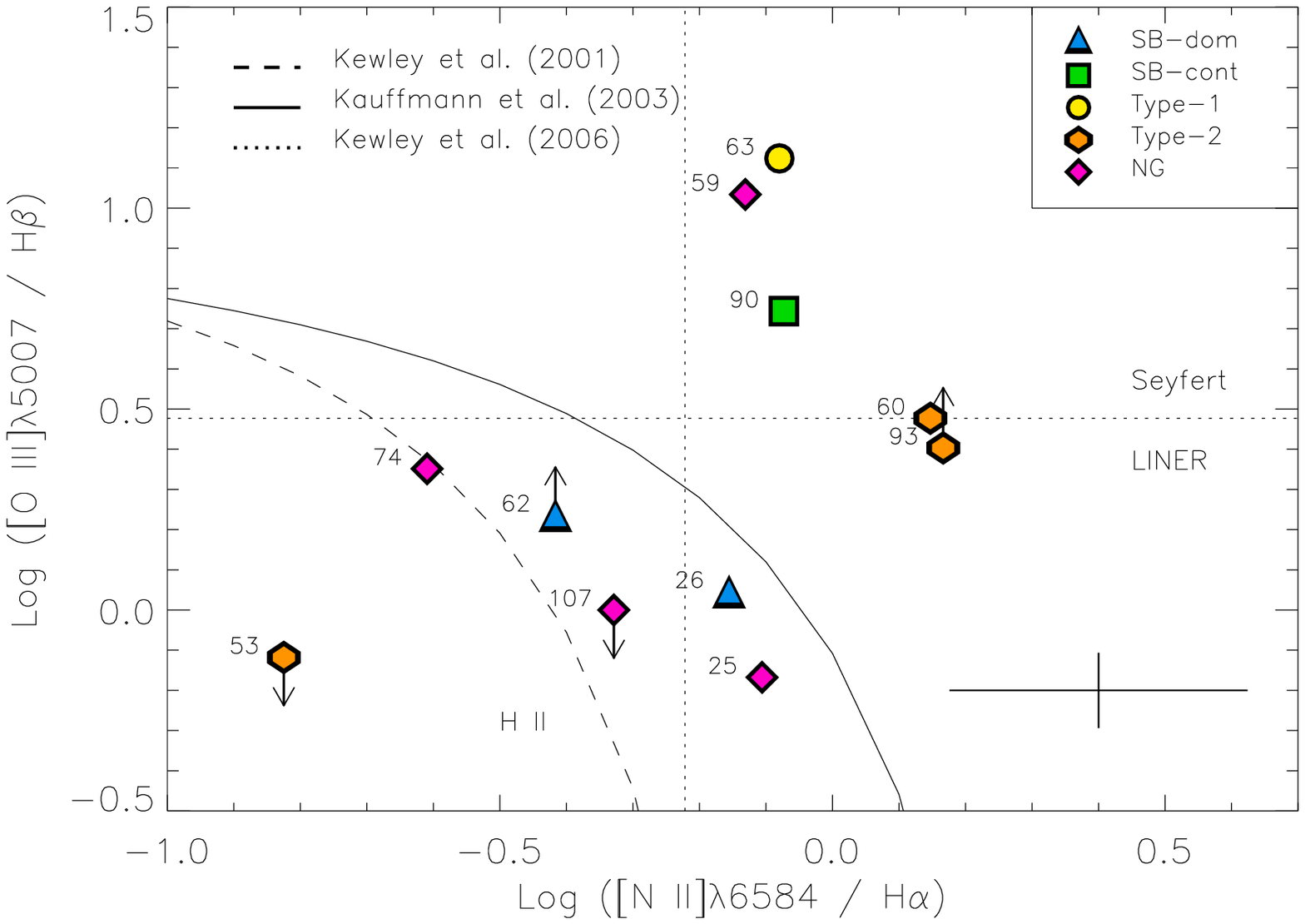}
\caption{BPT diagram involving the emission-line flux ratios $[O III]/H\beta$ and $[N II]/H\alpha$ 
for the 11 galaxies in our sample with detections of the four emission lines involved. The empirical separations between 
AGN and H II regions, as defined in \citet{Kewley01} and \citet{Kauffmann03},
are shown together with the boundaries between LINERs and Seyfert galaxies 
($[O III]/H\beta$=3 and $[N II]/H\alpha$=0.6). The error bars at the bottom right corner correspond 
to the typical uncertainties of the two ratios.}
\label{bpt}
\end{figure*}

To definitely confirm the agreement between the SED and spectroscopic classifications, 
we have plotted the line fluxes reported 
in Table \ref{nir1} in a Baldwin-Phillips-Terlevich (BPT) diagram \citep{Baldwin81}. In Figure 
\ref{bpt} we show the [O III]$\lambda5007/H\beta$ versus [N II]$\lambda6584/H\alpha$ ratios for the 
11 galaxies in our sample with 
the four emission lines involved detected. Despite the low number of galaxies represented in this BPT diagram, 
examples of the five SED groups are included. Note that we have only considered the fluxes of the narrow-line 
components in this figure. The empirical separations between AGN and H II regions (dashed line = \citealt{Kewley01};  
solid line = \citealt{Kauffmann03}) are shown together with the boundaries between LINERs and Seyfert galaxies 
($[O III]/H\beta$=3 and $[N II]/H\alpha$=0.6; \citealt{Kewley06}). 

As expected, the {\it Type-1} G63 and the {\it SB-cont} G90 are well above the H II region. 
The {\it Type-2}s G60 and G93 also lie in the AGN region
of the diagram, but with ratios more typical of the Seyfert/LINER transition zone. G60  
shows broad H$\alpha$ and H$\beta$ components in its spectrum, but appears much closer to the boundary 
between the H II and AGN regions than, for example, G63. The galaxy G107, which shows broad components 
in its recombination lines, although with FWHMs $\sim$ 900 km~s$^{-1}$ only, 
is located in the composite region between the \citet{Kauffmann03} and \citet{Kewley06} definitions. 
This agrees with our SED classification as a {\it NG} and with a possible star-forming origin of its
X-ray luminosity \citep{Pereira11,Ranalli12}, as discussed above. Thus,
G107 might be either a very weak AGN whose optical/IR emission is diluted by the host galaxy, or
a spiral galaxy with intense star formation (see Figure \ref{acs} and Table 1 in \citealt{Ramos09}).

Similar positions, also in the composite region, 
are occupied by the two {\it SB-dom} G26 and G62 and the {\it NG}s G25 and G74, 
confirming their SED classifications. Attending to their X-ray luminosities, only the galaxy G62 
(L$_X\leq$42.55 erg~s$^{-1}$) could not be an AGN.

The position of the sources G53 and G59 in the BPT diagram is noticeable. We classified G53 as 
a {\it Type-2}, and its L$_X>L_X^{AGN}$, but it is well below the H II boundary in the more 
restrictive \citet{Kewley01} definition. This source is likely a weak AGN with intense star formation. 
On the other hand, G59, whose SED was fitted with a spiral template in \citealt{Ramos09}
and has L$_X<L_X^{AGN}$, 
appears in the AGN region of the diagram, very close to the quasar G63. Thus, the galaxy G59 
is likely a heavily obscured Type-2 AGN. 
In summary, according to the emission line fluxes represented in Figure \ref{bpt}, our
SED classification is correct for 9/11 galaxies. 

Our previous work was aimed to distinguish between pure and host galaxy-dominated AGNs using SED fitting over 
a broad wavelength range. The use of accurate multiwavelength photometry, which 
is now available for several deep fields, to classify AGN, instead of spectroscopy, represents huge
savings of both telescope time and data reduction effort. From the comparison presented here, we find that the SED 
classification method employed in \citealt{Ramos09} works for 89\% of a representative sample of X-ray and MIR
sources at $z\sim0.8$ (see Table \ref{comparison_broad}). 

\begin{figure*}
\centering
\includegraphics[width=10cm]{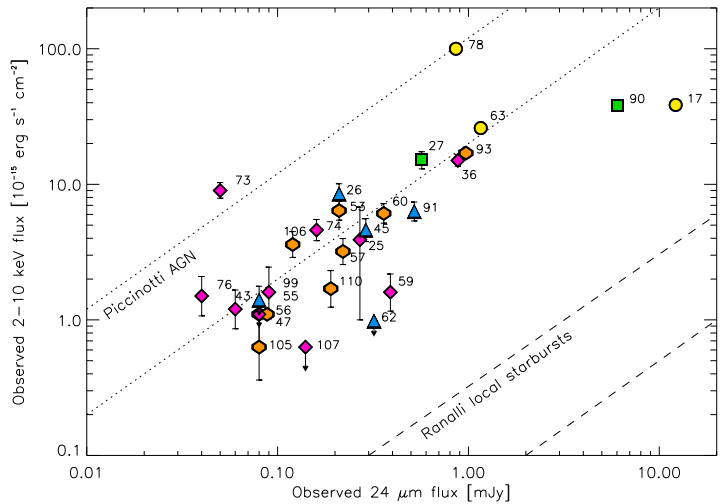}
\caption{Observed 24 \micron~versus 2-10 keV fluxes. The areas between dotted and dashed
lines correspond to the AGN region defined in \citet{Alonso04} from the \citet{Piccinotti82} X-ray-selected AGN 
sample and to the local starburst region as defined in \citet{Ranalli03} respectively. Symbols are the same
as in Figure \ref{bpt}. 24 \micron~flux errors are smaller than symbols (representative error is 0.01 mJy.}
\label{alonso}
\end{figure*}

To further investigate the level of AGN-dominance in these galaxies, in Figure \ref{alonso} we compare their
observed hard X-ray fluxes \citep{Waskett04,Nandra05} and 24 \micron~fluxes \citep{Barmby06} as in 
\citet{Alonso04}. The area between dotted 
lines corresponds to the extrapolation of the median hard X-ray-to-MIR ratios ($\pm 1\sigma$) 
of local AGN ($z<0.12$) selected in hard X-rays by \citet{Piccinotti82}. Same but for local starburst galaxies 
from \citet{Ranalli03} is the region between 
dashed lines. The plot looks very similar to Figure 1 in \citet{Alonso04}, where galaxies with redshifts in the range
z=0.2-1.6 were represented.
Some of the AGN occupy the local AGN region 
and the rest lie in the transition zone between local starbursts and AGN. Of the 28 objects in Figure 
\ref{alonso}, 18 are lying 
below the local AGN region ($\sim$64\%), indicating either high absorption in the X-rays or 
intense star formation contributing to the 24 \micron~emission. Considering the relatively low values of the hydrogen 
column densities reported in Table \ref{ha_lum} from \citet{Georgakakis06}, the first posibility is unlikely. 
We note, however, that the overall majority of the galaxies are very close to the local AGN region. 

The position of the galaxies G17 and G90 in Figure \ref{alonso} is remarkable. They are the 
brightest 24 \micron~sources in the whole sample studied in \citet{Barmby06}\footnote{13.72$\pm$0.01 and 
14.51$\pm$0.01 mag at 24 $\micron$ for G17 and G90 respectively.} and despite their strong H$\beta$ and H$\alpha$ broad 
components, they lie well below the AGN area of the diagram, indicating a strong contribution from 
star-heated dust at 24 \micron.

It is also worth mentioning the position of the galaxies G53 and G59 in Figure \ref{alonso}. Whereas G59 lies in 
the AGN region of the BPT diagram shown in Figure \ref{bpt}, in Figure \ref{alonso} appears considerably shifted 
from the local AGN region. Considering its low hydrogen column density (N$_H<0.58\times10^{22}~cm^{-2}$; \citealt{Georgakakis06}), 
the classification of this galaxy is unclear, althougth the BPT diagram confirms the 
presence of an active nucleus. 
Exactly the opposite happens to G53, which is definitely in the H II 
region of the BPT, but has typical MIR and hard X-ray AGN fluxes. In Table \ref{comparison_broad} we 
considered G53 and G59 as failures in the SED classification, but according to Figure \ref{alonso}, 
it appears to be correct. 

Summarising, in view of all the discussed diagnostic diagrams, X-ray luminosities and SED classification, 
we cannot confirm the presence of nuclear activity in 4/28 sources (14\% of the sample): G47, G55, G62 and G107. 
These galaxies 1) have L$_X<L_X^{AGN}$, 
%%{\bf 2) n$_H<5.8\times 10^{22}$ cm$^{-2}$ --indicative of relatively 
%%low obscuration\footnote{Heavily obscured AGN have typical hydrogen column densities of n$_H>10^{23-24}$ cm$^{-2}$; 
%%\citep{Risaliti99,Bauer10}}--,} 
2) SEDs fitted with galaxy or starburst templates 
and 3) lie outside the AGN region in Figures \ref{bpt} and \ref{alonso}. Note that they are the only four galaxies
in the sample with upper limits for their X-ray luminosities (see Table \ref{photoz}). They could be low-luminosity 
AGN heavily obscured in the X-rays, but we cannot confirm it. Thus, in the following, we 
will exclude these four galaxies from the results found for AGN, although, for simplicity, we will refer
to the whole sample of 28 galaxies as ``the AGN sample''.

\subsection{SFR from H$\alpha$ luminosity}
\label{ha_sfr}

By putting together the optical and NIR spectra of the galaxies, we have emission line 
flux measurements of H$\alpha$ for 18/28 galaxies and upper limits for another five (see Table \ref{nir1}).  
In Table \ref{ha_lum} we report the observed H$\alpha$ luminosities (L$_{H\alpha}$) obtained 
using the corresponding luminosity distances. 
In the absence of an AGN, the H$\alpha$ luminosity is the best indicator of the instantaneous 
SFR, since H$\alpha$ has little dependence on metallicity 
and it is much less affected by dust attenuation than the rest-frame UV continuum and Ly$\alpha$
\citep{Kennicutt98,Bicker05}.

In order to obtain SFRs for the galaxies in the sample using their H$\alpha$ luminosities,  
we have to correct L$_{H\alpha}$ from attenuation, using the A$_V$ values reported in Table \ref{ha_lum}. 
We calculated them using the H$\alpha$ and H$\beta$ narrow-line fluxes, when available, and 
the standard Galactic extinction curve of \citet{Cardelli89} with R$_V$=3.1:  \\

A$_V (mag) = 3.1 \times E(B-V) = 6.169 \times [Log (H\alpha/H\beta) - 0.486]$   \\

Then, we obtained reddening values at the wavelength of emission of H$\alpha$ (A$_{H\alpha}$), and 
we used them to correct L$_{H\alpha}$ from extinction.
In Table \ref{ha_lum} we report the attenuation-corrected H$\alpha$ luminosities (L$_{H\alpha}^{corr}$)
and corresponding errors.

\begin{figure*}
\centering
\includegraphics[width=10cm]{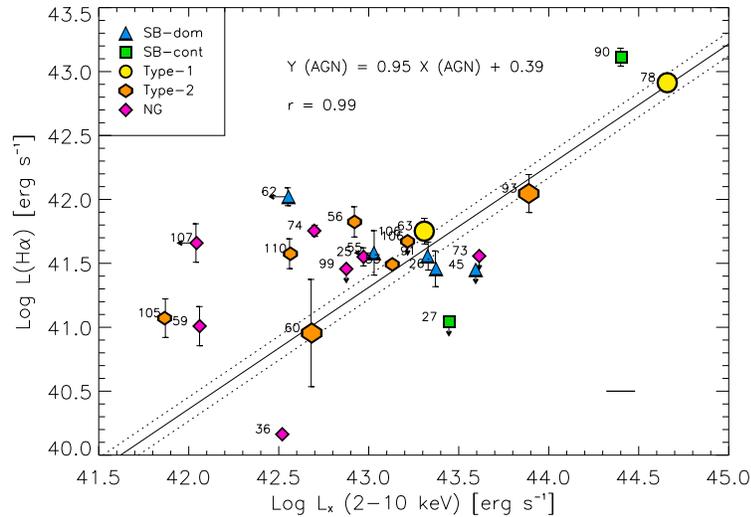}
\caption{H$\alpha$ versus hard X-ray luminosity (2-10 keV). 
Solid and dotted lines correspond to the AGN empirical relationship determined using the four  
AGN-dominated galaxies G60, G63, G78, and G93 (plotted with larger symbols) and 1$\sigma$ limits, respectively.
The correlation coefficient is r=0.99. 
Objects well above this line are likely dominated by star formation, whereas
those below may have H$\alpha$ fluxes affected by extinction. L$_X$ representative error is represented at 
the bottom right corner.}
\label{ha_xrays}
\end{figure*}

The H$\alpha$ emission in non-active galaxies is produced almost entirely by massive stars 
($M>10M_{\sun}$), but in active galaxies it includes a contribution from gas photoionised by the AGN. 
In fact, this contribution will dominate in the case of pure AGN, and it will be, in principle, less important 
in the case of buried AGN. To test the latter, in Figure \ref{ha_xrays} 
we represent L$_{H\alpha}$ versus L$_{2-10 keV}$ from \citet{Waskett04} 
and \citet{Nandra05}. Those X-ray luminosities are not absorption-corrected, but considering the relatively low values 
of the hydrogen column densities reported in Table \ref{ha_lum} from \citet{Georgakakis06}, we assume that they
are not going to differ significantly from the intrinsic hard X-ray luminosities. The H$\alpha$ luminosities
plotted in Figure \ref{ha_xrays} are not extinction-corrected either, because we only have attenuation-corrected 
values for half of the sample (see Table \ref{ha_lum}).

If we consider the 2-10 keV luminosity as a proxy of the AGN, it should be correlated with L$_{H\alpha}$
for AGN-dominated objects. Thus, we can derive an empirical relationship between L$_{H\alpha}$ and L$_X$ using the 
four galaxies that we can definitely classify as AGN-dominated from their spectra, SED fits, and 
diagnostic diagrams presented in this work: G60, G63, G78, and G93. By fitting them, we find
a correlation in the form Log (L$_{H\alpha}^{AGN}$) = 0.95 Log (L$_X$) + 0.39, with a correlation coefficient r=0.99.
G60 has a larger error bar than the other three pure-AGN (see Figure \ref{ha_xrays}) because of the multi-component 
fit performed to derive 
the H$\alpha$ flux, in addition to its lower H$\alpha$ emission as compared to that of G63, G78, and G93. 
However, in spite of its large error bar, G60 should be included it in the AGN fit because it is the only 
pure-AGN representative of the lower luminosity objects in the sample. To ensure that we are not introducing 
any bias in the determination of L$_{H\alpha}^{AGN}$ by including G60, we performed the test of excluding it from 
the AGN fit, and we found Log (L$_{H\alpha}^{AGN}$) = 0.87 Log (L$_X$) + 3.80, with r=0.98. The differences between 
the L$_{H\alpha}^{AGN}$ values derived from the two fits are much smaller 
than the L$_{H\alpha}^{SF}$ 
errors quoted in Table \ref{ha_lum}. Therefore, we can include G60 in the fit without introducing any
bias in the final results.

Of the four pure-AGN, G93 is the only one whose H$\alpha$ flux could be 
marginally affected by sky line contamination. Looking carefully at the exact position of the sky lines, 
none of them coincides with $\lambda$(H$\alpha$), but the residuals of the sky subtraction could
marginally affect the blue wing of H$\alpha$.
To test how this could affect the determination of L$_{H\alpha}^{AGN}$, we repeated the same exercise 
as for G60, but excluding G93. In this case we obtained Log (L$_{H\alpha}^{AGN}$) = 
0.97 Log (L$_X$) - 0.40, with r=0.99, which is even more similar to the original fit than when we excluded 
G60. Thus, including G93 makes no significant difference to L$_{H\alpha}^{AGN}$, 
and ultimately, to L$_{H\alpha}^{SF}$.

The luminosity of H$\alpha$ will be partly due to the AGN and to the star formation (L$_{H\alpha}$ = 
L$_{H\alpha}^{AGN}$ + L$_{H\alpha}^{SF}$). If the 2-10 keV luminosity probes the AGN, 
objects lying above the correlation will have a significant contribution from star formation in their H$\alpha$ 
emission (e.g. G56, G59, G62, G74, G105, G107 and G110). These galaxies have relatively low hard X-ray luminosities, 
indicating the presence of either a weak/absent AGN with intense star formation or a heavily obscured AGN. 
At the high luminosity end, the {\it SB-cont} 
G90 shows an excess of H$\alpha$ emission as well, explaining its position in Figure \ref{alonso}.
On the other hand, we have two galaxies lying below the correlation (e.g. G27 and G36). In the case of G36,
high attenuation could be affecting its H$\alpha$ emission, whereas in the case of G27 (A$_V\leq$0.23 mag) its H$\alpha$ flux could be
compromised by the line position, close to the edge of the detector (see Figure \ref{groth5}).

%It is noticeable the position of the three {\it SB-dom}s in Figure \ref{ha_xrays}, lying below the 
%AGN correlation. Based on their SED classification, we expect a higher value of L$_{H\alpha}$ due to  
%star formation. This could be due to the lack of extinction-correction of the H$\alpha$ 
%luminosities. 
%For example, in the case of the galaxy G26, A$_V$=3.53 mag. 
%If the H$\alpha$ emission of these galaxies is obscured, both the AGN and the star formation emission 
%should be revealed in the MIR, as we discuss in Section \ref{discussion}.

\begin{table*}
\centering
\scriptsize
\begin{tabular}{lccccccccccc}
%\rotate
%\tabletypesize{\tiny}
%\tablecaption{Luminosities and SFRs.}
\hline
\hline
ID & \multicolumn{1}{c}{Dist} & A$_V$ & \multicolumn{1}{c}{L$_{H\alpha}\times10^{41}$}& \multicolumn{1}{c}{L$_{H\alpha}^{corr}\times10^{41}$}&
\multicolumn{1}{c}{L$_{H\alpha}^{SF}\times10^{41}$} & SFR (L$_{H\alpha}^{SF}$) & F$_{24}^{obs}$ &SFR (F$_{24}^{SF}$)  & L$_X\times10^{43}$ & N$_H\times10^{22}$ & Group \\
 & (Mpc) &  (mag)& (erg~s$^{-1}$)& (erg~s$^{-1}$) & (erg~s$^{-1}$) & (M$_{\sun}$~yr$^{-1}$)& (mJy) & (M$_{\sun}$~yr$^{-1}$)  & (erg~s$^{-1}$) & (cm$^{-2}$) & \\
\hline
 17  & 8597 &  \dots	       & \dots  	  &   \dots          &   \dots         & \dots        & 12.19$\pm{0.11}$ (74\%) &$>$1000  & 33.9$\pm$2.5    &  \dots  & 3 \\
 25  & 4481 &  1.77$\pm{1.11}$ & 3.55$\pm{0.58}$  & 13.3$\pm{11.3}$  &   11.4$\pm$11.3 & 6$\pm$6      &  0.27$\pm{0.01}$ (69\%) & 8	  & 0.93$\pm$0.69   &  \dots  & 5 \\
 26  & 4823 &  3.53$\pm{2.40}$ & 2.86$\pm{0.92}$  & 40.0$\pm{72.9}$  &   35.4$\pm$72.9 & 19$\pm$40    &  0.21$\pm{0.01}$ (35\%) & 17      & 2.35$\pm$0.44   &  \dots  & 1 \\
 27  & 3916 &  $\leq$0.23      & $\leq1.10$       &$\leq$1.10        &   $\leq$1.10    & $\leq$0.6    &  0.57$\pm{0.01}$ (85\%) & 6	  & 2.77$\pm$0.40   &  \dots  & 2 \\
\hline  			 
 36  & 1358 &  \dots	       & 0.15$\pm{0.01}$  &$\geq$0.15        &   \dots         & \dots        &  0.88$\pm{0.01}$ (36\%) & 4	  & 0.33$\pm$0.03   &    8.82 & 5 \\
 43  & 2885 &  \dots	       & \dots  	  &   \dots          &   \dots         & \dots        &  0.06$\pm{0.01}$ ( 0\%) & 2	  & 0.12$\pm$0.04   &    3.36 & 5 \\
 45  & 8449 &  \dots	       & $\leq2.82$       &   \dots          &   \dots         & \dots        &  0.29$\pm{0.01}$ ( 5\%) & 1000    & 3.91$\pm$0.76   & $<$0.08 & 1 \\
 47  & 2159 &  \dots	       & \dots            &   \dots          &   \dots         & \dots        &  0.08$\pm{0.01}$ ( 7\%) & 1	  & $\leq$0.06      & $<$0.08 & 5 \\
 53  & 4203 &  2.21$\pm{0.99}$ & 3.11$\pm{0.23}$  & 16.2$\pm{12.0}$  &   13.4$\pm$12.0 & 7$\pm$7      &  0.21$\pm{0.01}$ (72\%) & 4	  & 1.35$\pm$0.22   &    5.51 & 4 \\
 55  & 7990 &  \dots	       & 3.82$\pm{1.53}$  &$\geq$3.82        &$\geq$1.63       & $\geq$1      &  0.08$\pm{0.01}$ ( 4\%) & 77      & $\leq$1.06      & $<$2.15 & 1 \\
 56  & 7963 &  $\geq$0.97      & 6.67$\pm{1.82}$  &$\geq$13.8        &$\geq$12.0       & $\geq$7      &  0.08$\pm{0.01}$ (82\%) & 4	  & 0.83$\pm$0.42   &    0.53 & 4 \\
 57  & 3681 &  \dots	       & \dots  	  &   \dots	     &   \dots         & \dots        &  0.22$\pm{0.01}$ (51\%) & 7       & 0.52$\pm$0.11   &    0.33 & 4 \\
\hline  			       
 59  & 2451 &  0.99$\pm{2.21}$ & 1.02$\pm{0.36}$  & 2.14$\pm{3.61}$  &   1.88$\pm$3.61 & 1$\pm$2      &  0.39$\pm{0.01}$ (81\%) & 2	  & 0.11$\pm$0.04   & $<$0.58 & 5 \\
 60  & 2571 &  0.24$\pm{5.96}$ & 0.90$\pm{0.87}$  & 1.08$\pm{4.91}$  &   0.05$\pm$4.92 &0.03$\pm$2.71 &  0.36$\pm{0.01}$ (41\%) & 7	  & 0.48$\pm$0.08   & $<$0.24 & 4 \\
 62  & 5528 &  \dots	       & 10.5$\pm{1.7}$   &$\geq$10.5        &$\geq$9.72       & $\geq$5      &  0.32$\pm{0.01}$ ( 0\%) & 94      & $\leq$0.36      & $<$5.84 & 1 \\
 63  & 2558 &  7.04$\pm{2.30}$ & 5.64$\pm{1.30}$  & 1084$\pm{1880}$  &   1080$\pm$1880 & 595$\pm$1034 &  1.16$\pm{0.01}$ (100\%)& \dots   & 2.02$\pm$0.15   & $<$0.12 & 3 \\
 73  & 6179 &  \dots	       & $\leq3.61$       &   \dots          &   \dots         & \dots        &  0.05$\pm{0.01}$ ( 0\%) & 9	  & 4.09$\pm$0.55   &    0.72 & 5 \\
 74  & 3009 &  1.39$\pm{1.05}$ & 5.68$\pm{0.57}$  & 16.1$\pm{12.7}$  &   15.0$\pm$12.7 & 8$\pm$7      &  0.16$\pm{0.01}$ (25\%) & 5	  & 0.50$\pm$0.09   &    2.16 & 5 \\
 76  & 1303 &  \dots	       & \dots  	  &   \dots	     &   \dots         & \dots        &  0.04$\pm{0.01}$ (23\%) & 0.1     & 0.03$\pm$0.01   &    0.02 & 5 \\
 78  & 6171 &  \dots	       & 81.8$\pm{0.6}$	  &$\geq$81.8        &$\geq$4.12       &$\geq$2       &  0.86$\pm{0.01}$ (97\%) & 3	  & 45.3$\pm$1.7    & $<$0.04 & 3 \\
\hline  			     
 90  & 7474 &  2.41$\pm{3.34}$ & 130$\pm{21}$     & 784$\pm{1961}$   &   740$\pm$1961  & 407$\pm$1079 &  6.02$\pm{0.05}$ (91\%) &$>$1000  & 25.3$\pm$1.6    &   1.16  & 2 \\
 91  & 5307 & \dots	       & 3.60$\pm{0.91}$  &$\geq$3.60        &   \dots         & \dots        &  0.52$\pm{0.01}$ ( 0\%) & 163     & 2.11$\pm$0.34   &  20.77  & 1 \\
 93  & 6173 & \dots	       & 11.1$\pm{3.8}$   &$\geq$11.1        &   \dots         & \dots        &  0.97$\pm{0.01}$ (65\%) & 166     & 7.71$\pm$0.73   &   3.89  & 4 \\
 99  & 6262 & \dots	       & $\leq2.86$       &   \dots          &   \dots         & \dots        &  0.09$\pm{0.01}$ (20\%) & 17      & 0.75$\pm$0.33   &   3.63  & 5 \\
105  & 3123 & \dots	       & 1.18$\pm{0.41}$  &$\geq$1.18        &$\geq$1.01       &$\geq$0.6     &  0.08$\pm{0.01}$ (69\%) & 1	  & 0.07$\pm$0.04   &   2.25  & 4 \\
106  & 6179 & \dots	       & $\leq4.70$       &   \dots          &   \dots         & \dots        &  0.12$\pm{0.01}$ (74\%) & 4	  & 1.64$\pm$0.36   &  20.46  & 4 \\  
107  & 3830 &  2.22$\pm{2.43}$ & 4.56$\pm{1.58}$  & 23.9$\pm{44.2}$  &$\geq$23.7       & $\geq$13     &  0.14$\pm{0.01}$ (30\%) & 7	  & $\leq$0.11      & $<$2.12 & 5 \\
110  & 4250 & \dots	       & 3.76$\pm{1.01}$  &$\geq$3.76        &$\geq$2.97       & $\geq$2      &  0.19$\pm{0.01}$ (96\%) & 0.2     & 0.37$\pm$0.11   & $<$0.26 & 4 \\
\hline	
\end{tabular}				 					
\caption{Observed, attenuation-corrected, 
and AGN-corrected H$\alpha$ luminosities calculated using the fluxes reported in 
Table \ref{nir1} and the luminosity distances listed in column 2. The optical extinction listed in column 3 has been obtained
from the observed H$\alpha$/H$\beta$ ratios when available. Columns 7 lists the SFRs obtained from L$_{H\alpha}^{SF}$. 
Columns 8 corresponds to the observed 24 \micron~fluxes and the percentage of AGN contamination removed from the 
latter to calculate the IR SFRs reported in column 9, which have an accuracy of better than 0.2 dex; \citep{Rieke09}. Columns 10, 11, and 12 indicate the 
2-10 keV X-ray luminosities from \citet{Nandra05} and \citet{Waskett04}, the column densities from \citet{Georgakakis06}, and the SED
classification as in Table \ref{photoz}.}
\label{ha_lum} 
\end{table*}

Galaxies lying close to 
the AGN correlation shown in Figure \ref{ha_xrays} will have L$_{H\alpha}\approx L_{H\alpha}^{AGN}$, 
whereas for those located well above, the L$_{H\alpha}^{SF}$ contribution to H$\alpha$ will be significant. 
Thus, we use our AGN empirical correlation to 
calculate, for a given L$_X$, the expected L$_{H\alpha}^{AGN}$ of the galaxies. We then subtract
this AGN contribution from the attenuation-corrected L$_{H\alpha}^{corr}$ values to obtain L$_{H\alpha}^{SF}$ (see 
Table \ref{ha_lum}). For the galaxies well below the correlation (L$_{H\alpha} << L_{H\alpha}^{AGN}$) we 
cannot estimate the AGN contribution to H$\alpha$ even when L$_{H\alpha}$ has been extinction-corrected, and 
thus we have used the total L$_{H\alpha}$ values as upper limits.

Ideally, this evaluation of the AGN contribution should have been done using attenuation-corrected 
H$\alpha$ luminosities
and intrinsic hard X-ray fluxes, since reddening in the optical and in the X-rays can be different \citep{Fiore12}. 
However, we only have A$_V$ values for 9 galaxies and upper limits for another two. Besides, out of the four 
AGN-dominated sources that we use to derive the AGN correlation, only G60 has a reliable extinction 
value\footnote{We have likely overestimated the A$_V$ value for G63 (A$_V$=7$\pm$2 mag). The fitting of its 
H$\beta$ profile using multiple components was particularly difficult.}. On the other
hand, the observed and intrinsic hard X-ray luminosities will be similar, considering the
low N$_H$ values \citep{Georgakakis06}. Thus, our method
will be more reliable for galaxies with similar levels of obscuration in the optical and in the X-rays
and vice versa.

We can now estimate SFRs for the 
galaxies in our sample using our individual L$_{H\alpha}^{SF}$ values and the 
\citet{Kennicutt98} conversion for Case B recombination at T$_e$=10,000 K \citep{Osterbrock89}, assuming solar 
metallicity and a \citet{Kroupa03} initial mass function (IMF):

SFR (M$_{\sun}~yr^{-1}) = 5.5\times10^{-42} L_{H\alpha}^{SF}~(erg~s^{-1})$

%For the galaxies lying well above the empirical AGN relationship 
We obtain SFR(L$_{H\alpha}^{SF}$) = [0.03, 19] M$_{\sun}~yr^{-1}$, and two extreme 
values of 407 and 595 ($\pm$1000) M$_{\sun}~yr^{-1}$ for the {\it Type-1}s G63 and G90 respectively. 
In the case of G63, it is likely that we have overestimated the extinction (A$_V$=7$\pm$2 mag), 
because the fit of the H$\beta$ profile using multiple components was particularly challenging. 
This galaxy does not appear to have such an intense star formation in Figure \ref{alonso}, as 
opposed to G90. We will discuss the case of these two galaxies in Section \ref{mips_sfr}.

The individual SFRs, with the exception of those of G63 and G90, are among the lowest  
reported in the literature for samples of non-active star-forming galaxies at 
similar redshifts and stellar masses (e.g. \citealt{Twite11} and Rodr\' iguez-Eugenio et al. in prep.), 
as we discuss in Section \ref{discussion}.

The average and median SFRs, measured using the values reported in Table \ref{ha_lum}, including 
upper and lower limits as fixed values, are 5$\pm$5 and 2 M$_{\sun}~yr^{-1}$. We have excluded
the extreme SFRs of G63 and G90 because of their 
large uncertainties, and those of G55, G62 and G107 because they are not confirmed AGN.

If we do not consider the upper and lower limits (i.e., if we only use the SFRs of the 
galaxies G25, G26, G53, G59, G60 and G74) the average and median SFRs are 7$\pm$7 and 7 M$_{\sun}~yr^{-1}$.
Note that none of these six galaxies has its H$\alpha$ emission affected by sky line contamination.

\subsection{SFR from observed 24 \micron~emission}
\label{mips_sfr}

In dusty galaxies, the UV light emitted by young stars is absorbed by dust and re-emitted in the IR. 
Therefore, the IR luminosity, either monochromatic or total, 
can be used to estimate the SFR \citep{Kennicutt09}. In particular, the 24 
\micron~luminosity is a good indicator of
the current SFR of dusty star-forming galaxies (e.g. \citealt{Alonso06,Calzetti07,Rieke09}).

\begin{figure*}
\centering
\par{
\includegraphics[width=8.0cm]{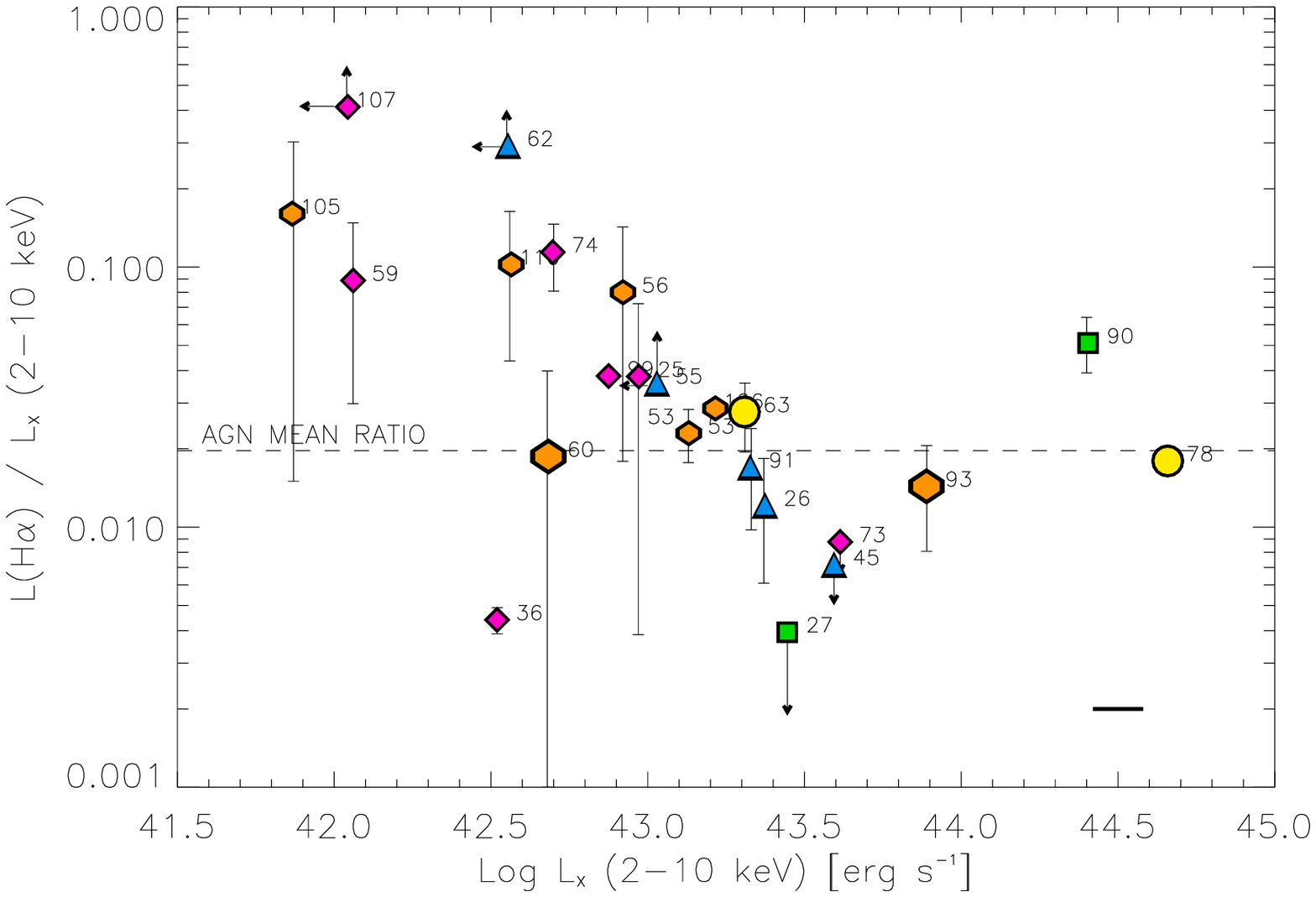}
\includegraphics[width=8.0cm]{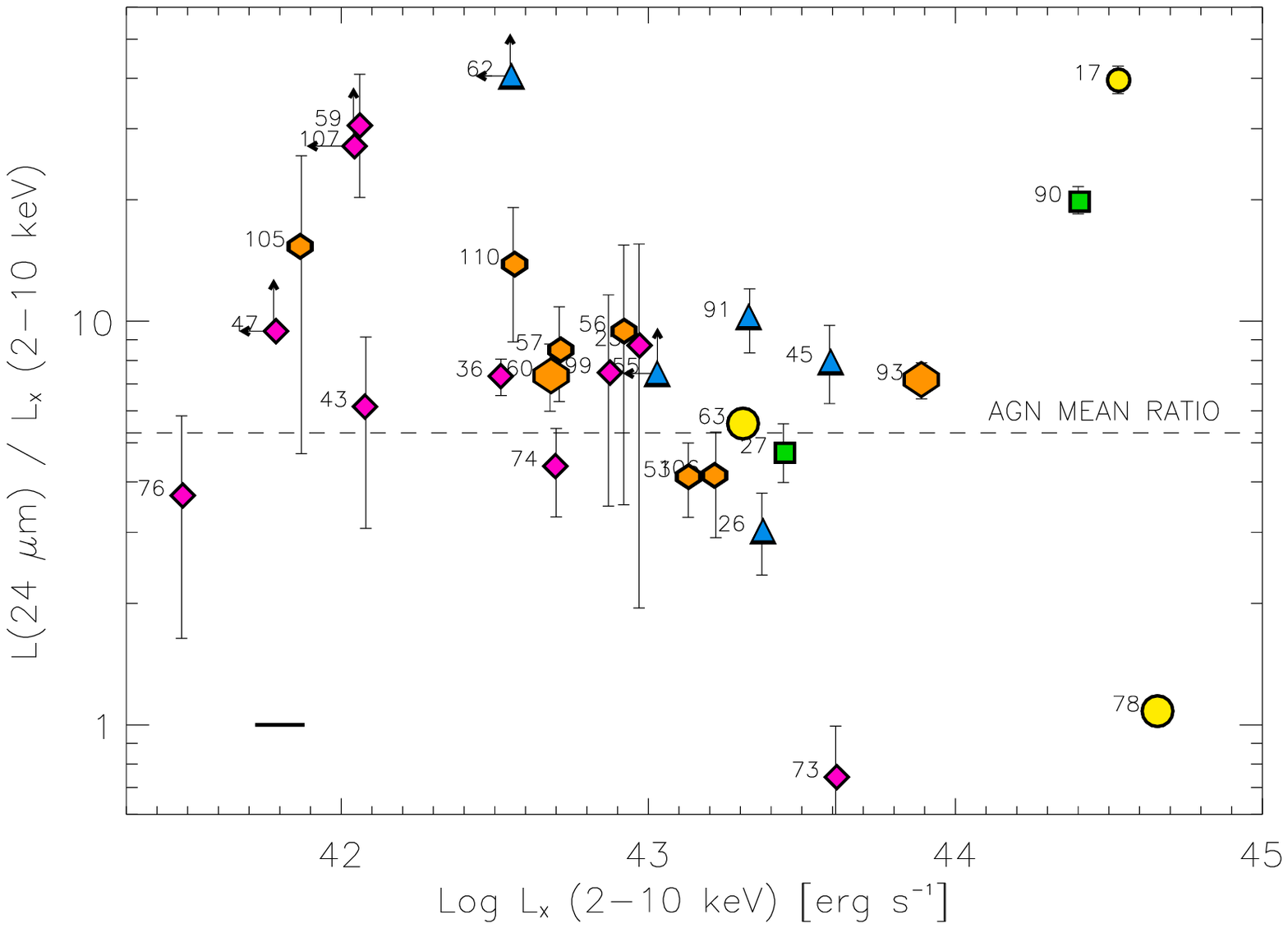}\par}
\caption{Left: L$_{H\alpha}$/L$_X$ versus hard X-ray luminosity (2-10 keV). Objects with 
a significant contribution of H$\alpha$ due to star formation are 
clearly offset from the locus of galaxies. The horizontal dashed line corresponds to the mean 
value of L$_{H\alpha}$/L$_X$, considering the four pure AGN G60, G63, G78, and G93. 
Right: same as in the left panel, but for L$_{24}$/L$_X$. L$_X$ representative error is represented at 
the bottom of the two panels.}
\label{ha_mips_xrays}
\end{figure*}

For galaxies whose H$\alpha$ emission is obscured, 
both the AGN- and star-heated dust should be detected in the MIR. Thus, by comparing
the H$\alpha$ and 24 \micron~luminosities we can distinguish between galaxies with apparently low SFRs due to high levels
of extinction and galaxies with negligible star formation. In Figure \ref{ha_mips_xrays} we
compare the individual values of L$_{H\alpha}$/L$_X$ and L$_{24}$/L$_X$. The horizontal lines
represent the mean AGN ratio in each case, using the four pure-AGN G60, G63, G78, and G93. 
The galaxies G56, G59, G90, G105, G107, and G110 are well above the mean AGN ratios, 
indicating a strong contribution of star formation.

On the contrary, the {\it NG} G74, which 
in the left panel of Figure 
\ref{ha_mips_xrays} is clearly above the AGN mean L$_{H\alpha}$/L$_X$, has a low L$_{24}$/L$_X$ 
ratio. As discussed in \citet{Kennicutt09}, the galaxies with relatively low H$\alpha$ luminosities (L(H$_{\alpha}$)$<10^{42}$
 erg~s$^{-1}$) tend to have little dust, and consequently low A$_V$ and weak IR emission. This could be
the case of the galaxies G53 and G74. 

The four galaxies for which we cannot confirm the presence of nuclear activity, G47, G55, G62
and G107, show large L$_{24}$/L$_X$ values, indicative of warm dust emission produced by intense star formation and/or
a heavily obscured AGN.
The position of the galaxies G63 and G90 in the two panels of Figure \ref{ha_mips_xrays} confirms 
what we discussed in Section \ref{ha_sfr}. G63 does not have a large L$_{24}$/L$_X$ value as it is the 
case for G90. Thus, the extreme SFR measured from its H$\alpha$ luminosity is likely the result of 
an overestimation of the A$_V$, whereas in the case of G90 it is probably real, although affected by 
a large uncertainty. In addition, G90 has its H$\alpha$ emission somehow affected by sky line contamination.

The largest L$_{24}$/L$_X$ value corresponds to the {\it Type-1} G17, for which we do not have an
H$\alpha$ measurement due to the galaxy redshift ($z$=1.28). 
Finally, the {\it SB-dom} G45 and G91 show a
MIR excess as compared to the H$\alpha$ emission, indicative of relatively large A$_V$ and in agreement with their
SED classification as {\it SB-dom}.

To estimate SFRs using observed 24 \micron~emission, it is necessary 
to remove the contribution of AGN-heated dust from the total MIR emission, as we did for 
the H$\alpha$ luminosities. 
%The 24 \micron~fluxes of the galaxies in the sample will mainly come from the active nucleus for
%AGN-dominated sources (e.g. G63 and G78), and even for host galaxy-dominated objecs 
%(e.g. G62 and G74), there will be some AGN-heated dust contribution to the MIR emission. 

We have estimated the AGN contribution to the 24 \micron~emission of the individual 
galaxies using the SED fits shown in Figure \ref{seds}. For {\it Type-1} and {\it Type-2}, 
we have subtracted the 24 \micron~flux of the fitted template from the observed 24 \micron~flux.
We assume that the flux difference corresponds to dust heated by star formation. The median 
AGN contributions are 97\% in the case of {\it Type-1} and 72\% for {\it Type-2}
(see column 8 in Table \ref{ha_lum}).
For the galaxies classified as {\it SB-dom} and {\it NG}, we consider that 
the 24 \micron~fluxes of the fitted templates are representative of the star formation, and 
any excess in the observed fluxes is due to the AGN (median AGN contribution of 4\% in the case of 
{\it SB-dom} and 25\% for {\it NG}). Finally, 
for the two {\it SB-cont}, we have repeated the SED fits using a {\it Type-2} template 
in the case of G27, and a {\it Type-1} template for G90, and consider the difference between 
the fitted and observed 24 \micron~fluxes as due to star-heated dust. The AGN contributions 
are 85\% for G27 and 91\% for G90. 
 
We could have corrected the 24 \micron~fluxes from AGN contamination using the same technique 
employed in Section \ref{ha_sfr}, but then we would have been left with no correction for those
galaxies lying below the AGN mean ratio in Figure \ref{ha_mips_xrays}. 
%Thus, we prefer to estimate F$_{24}^{SF}$ values for the 28 galaxies in the sample using the
%SED fits, as described above. 

%The AGN-dominated galaxies in the sample will have a strong contribution from 
%AGN-heated dust at 24 \micron, in addition to the star formation, and 
%for those AGN that are either masked by the host galaxy emission, or intrinsically weak, the 
%AGN emission is more likely to be unveiled in the MIR. 
%Thus, in 
%both AGN- and host galaxy-dominated galaxies, the SFRs (L$_{24}$) will be overestimated because
%of the AGN contamination. On the other hand, the SFRs based on the H$\alpha^{corr}$ luminosities 
%of masked/weak AGN will be more reliable thanks to the AGN obscuration. In fact, they are very 
%similar to the SFRs reported for quiescent galaxies at similar redshift (e.g. \citealt{Doherty04}).   

Using the AGN-corrected 24 \micron~fluxes F$_{24}^{SF}$ and equation 14 in \citet{Rieke09}:

Log SFR (M$_{\sun}~yr^{-1}) = A(z) + B(z)[Log (4\pi D_L^2 F_{24}^{SF}) - 53]$,

%7.8\times10^{-10}~L_{24}(L_{\sun}) \times [7.76\times10^{-11} L_{24}(L_{\sun})]^{0.048}$

we can calculate IR SFRs for the 28 galaxies in our sample (see Table \ref{ha_lum}). 
A(z) and B(z) are the SFR fit coefficients 
as a function of redshift for MIPS, and the above calibration assumes the \citet{Kroupa03} IMF. 
We obtain values of SFR (F$_{24}^{SF}$) = [0.1, 166] M$_{\sun}~yr^{-1}$  and three extreme
values of $\ga$1000 M$_{\sun}~yr^{-1}$ for the galaxies G17, G45, and G90. 
SFRs larger than 1000 M$_{\sun}~yr^{-1}$ are only measured in 
ULIRGs and/or extreme starbursts \citep{Magnelli10}. Given the position of G17 and G90 in 
Figure \ref{alonso} and in the right panel of Figure \ref{ha_mips_xrays}, the
two galaxies must be ULIRGs. In fact, they both have monochromatic 24 \micron~luminosities
larger than 45.5 erg~s$^{-1}$, exclusive of this type of object \citep{Alonso07}. 
The case of G45 is different. In the right panel of Figure \ref{ha_mips_xrays}
the galaxy does not show a strong excess in 24 \micron~over the AGN mean ratio. This indicates that we may have
underestimated its AGN contribution to the 24 \micron~flux (only 5\%). 

In general we find SFR (F$_{24}^{SF}) \ga SFR~(L_{H\alpha}^{SF}$), although for some 
galaxies SFR (F$_{24}^{SF}) < SFR~(L_{H\alpha}^{SF}$). This is 
expected, since the relation between H$\alpha$ and 24 \micron~is nonlinear, 
even when the H$\alpha$ emission has been corrected from attenuation (see Figure 6 in \citealt{Kennicutt09}).
The average and median SFRs that we obtain from the 24 \micron~fluxes are
20$\pm$50 and 5 M$_{\sun}~yr^{-1}$ respectively. We have excluded the extreme values of 
G17, G90 and G45 --which we do not consider representative of the sample-- and those of the non-confirmed 
AGN. The median SFR is in agreement with that derived from the attenuation-corrected H$\alpha$ luminosities. 
This similarity between the H$\alpha$ and 24 \micron~median SFRs 
gives us extra-confidence in the AGN corrections that we have employed here.

\section{Comparison with samples of non-active galaxies}
\label{discussion}

In Sections \ref{ha_sfr} and \ref{mips_sfr} we have estimated SFRs from attenuation- and AGN-corrected 
H$\alpha$ luminosities and from AGN-corrected 24 \micron~fluxes. We obtain average SFRs = 7$\pm$7 
and 20$\pm$50 M$_{\sun}~yr^{-1}$ respectively (median SFRs = 7 and 5 M$_{\sun}~yr^{-1}$).
We now compare our SFRs with those published in the literature for different samples 
of non-active galaxies of similar stellar masses and redshifts, once converted to the \citet{Kroupa03} IMF. 
The results of this comparison, including the conversion factors, when necessary, are summarised in Table \ref{comparison}. 
Although we do not know the individual stellar masses of our galaxies, the sample is representative 
of the X-ray selected population of AGN at $z\sim0.8$, as we showed in Section \ref{sample}, which
typically have stellar masses in the range 0.8-1.2$\times 10^{11} M_{\sun}$ 
(e.g. \citealt{Alonso08}). 
%Thus, by comparing our SFRs with those measured for star-forming galaxies of similar stellar 
%masses, we find that the hosts of AGN at $z\sim0.8$ seem to be less efficient forming stars. 

We first compare with the small sample of 7 star-forming galaxies selected by \citet{Doherty04} from 
the \citet{Cohen00} magnitude-limited sample (R$<$23 mag) at $z\sim0.8$.
We used the H$\alpha$ fluxes and reddening values reported in Tables 1 and 2 in \citet{Doherty04}
to calculate attenuation-corrected SFRs exactly as we did for our AGN. We obtained 
SFRs=[4, 11] M$_{\sun}~yr^{-1}$ for the individual galaxies, and a mean of 7$\pm$3 M$_{\sun}~yr^{-1}$,
which is very similar to ours. Unfortunately, \citet{Doherty04} did
not give an indication of the stellar masses of their sample.

Secondly, Rodr\' iguez-Eugenio et al.~(in preparation) observed
a sample of 30 star forming galaxies at $z\sim1$ and with M$_*\sim10^{10.8}M_{\sun}$ in the EGS 
with LIRIS MOS, for which 
there are DEEP2 optical spectra as well. They find an average SFR from aperture- and attenuation-corrected
H$\alpha$ luminosities of 23$\pm$17 M$_{\sun}~yr^{-1}$, with individual values ranging from 5 to 64
M$_{\sun}~yr^{-1}$.

\citet{Twite11} also used LIRIS MOS observations of galaxies at $z\sim1$ and 
find attenuation-corrected values of the SFR = [4, 319]
M$_{\sun}~yr^{-1}$, with an average SFR of 66$\pm$83 M$_{\sun}~yr^{-1}$ for the 14 massive galaxies 
(M$_*>10^{10.5}M_{\sun}$) with extinction-corrected H$\alpha$ measurements at the 
1.5$\sigma$ detection level.

Finally, based on H$\alpha$ imaging of a sample of 153 star-forming galaxies at $z\sim0.8$ with 
typical stellar masses of $\sim10^{10}M_{\sun}$, \citet{Villar11} determined attenuation-corrected 
SFRs = [3, 23] M$_{\sun}~yr^{-1}$, with a median value of 8 M$_{\sun}~yr^{-1}$.
This median SFR is very similar to ours (see Table \ref{comparison}). However, 
the average stellar mass of the \citet{Villar11} sample is $\sim10^{10}M_{\sun}$, whereas the typical 
mass of AGN at $z\sim0.8$ is $\sim10^{11}M_{\sun}$ \citep{Alonso08}. 
As represented in Figure 14 in \citet{Villar11}, the more massive the galaxies, the larger the
SFRs \citep{Dutton10,Villar11}. Thus, it is expected that the galaxies studied 
in \citet{Villar11} have lower SFRs than those studied by \citet{Twite11}
and Rodr\' iguez-Eugenio et al.~(in preparation).

\begin{table*}
\begin{tabular}{lcccccc}
%\tablecaption{Comparison with SFRs (from H$\alpha$) of non-active star-forming galaxies from the literature}
\hline
\hline
Work & Redshift & M$_*$ ($M_{\sun}$) & SFRs & Average SFR & Median SFR & IMF factor \\
\hline
\citet{Doherty04}           & $\sim$0.8 & \dots          & [4, 11]   &  7$\pm$3     & 6		&  \dots	\\
Rodr\' iguez-Eugenio et al. & $\sim$1   & $10^{10.8}$    & [5, 64]   & 23$\pm$17    & 19    	&  0.696 (a)	\\
\citet{Twite11}             & $\sim$1   & $>10^{10.5}$   & [4, 319]  & 66$\pm$83    & 54	&  1.196 (b) 	\\
\citet{Villar11}   	    & $\sim$0.8 & $10^{10}$      & [3, 23]   & \dots        & 8         &  0.696 (a)	\\
This work                   & $\sim$0.8 & $\sim10^{11}$  & [0.03, 19]& 7$\pm$7      & 7         &  \dots	\\
\hline
\end{tabular}   					 					
\caption{Comparison with SFRs (from H$\alpha$) of non-active star-forming galaxies from the literature. 
Columns 2 and 3 list the average/median redshift and an estimation of the stellar mass
of the samples considered.In the case of our AGN sample, we considered a stellar mass of $\sim10^{11}~M_{\sun}$, typical of AGN
at this redshift \citep{Alonso08}. Columns 4, 5, and 6 give the intervals, average {\bf ($\pm$ standard deviation), }
and median SFRs of the different samples, once
converted to the \citet{Kroupa03} IMF.
Finally, column 7 lists the conversion factor applied to convert the SFRs to the \citet{Kroupa03} IMF. Refs: (a) 
\citet{Salpeter55}; (b) \citet{Chabrier03}.}
\label{comparison} 
\end{table*}

Again, in spite of the reduced size of the samples compared here, as well as the uncertainty 
affecting our AGN-corrected SFRs, the comparison presented might indicate
that the presence of an AGN in a galaxy of M$_*\sim10^{11} M_{\sun}$ at $z\sim0.8$, 
%independently of the dominance of the AGN over the host galaxy emission, 
would be quenching its star formation. 
This quenching would be reducing the SFR from 20-50 M$_{\sun}~yr^{-1}$ (typically found for samples of 
non-active star-forming galaxies at this redshift and stellar mass: \citealt{Noeske07,Twite11}; 
Rodr\' iguez-Eugenio et al.~in prep.) to less than 10 M$_{\sun}~yr^{-1}$.
Alternatively, we might be seeing a delay between the offset of the star formation and AGN activity, 
as observed in the local universe \citep{Davies07,Wild10}.

\section{Comparison with other samples of active galaxies}
\label{discussion2}

Star formation activity in the hosts of AGN at intermediate redshift have been studied at longer wavelenghts than those
analysed here. For example, in a recent work based on FIR data from
the Herschel Space Observatory, \citet{Santini12} reported evidence of a higher average star formation activity 
in the hosts of X-ray selected AGN at $0.5<z<2.5$ compared to a mass-matched control sample of inactive galaxies.
This enhancement is found to be higher for the most luminous AGN in the sample. However, when they only consider 
star-forming galaxies in the control sample, they found roughly 
the same level of star formation activity in the hosts of AGN and non-active galaxies. 

A similar result is reported by 
\citet{Lutz10}, but based on submillimeter data of a sample of 895 X-ray selected sources in the Chandra 
Deep Field South (CDFS) and Extended-CDFS (ECDFS). The latter authors analysed stacked emission at 870 \micron, 
representative of X-ray-selected AGN at $z\sim$1, and inferred an average SFR$\sim$31 M$_{\sun}~yr^{-1}$, once
converted to the \citet{Kroupa03} IMF, and assuming star formation-dominated submillimeter emission. 
As claimed by \citet{Lutz10}, 30 M$_{\sun}~yr^{-1}$ is among the typical values found for samples of 
non-active star-forming galaxies at $z\sim$1 and M$_* \ga 10^{10.5}M_{\sun}$ (e.g. \citealt{Noeske07}). 
In fact, this value is very similar to the average SFRs reported by Rodr\' iguez-Eugenio et al.~(in prep.)
and \citet{Twite11} for samples of star forming galaxies at the same redshift and within the same range
of stellar masses (see Table \ref{comparison}).

The star formation properties of 58 X-ray-selected AGN at $0.5<z<1.4$ were studied by \citet{Alonso08}
by modelling their multifrequency SEDs. As in the previously mentioned works, they do not find strong 
evidence in the host galaxies of those AGN for either highly suppressed or enhanced star formation 
when compared to a mass-matched sample of galaxies at the same redshifts. However, these AGN were
selected to have SEDs dominated by stellar emission, and thus, they are representative of only 50\%
of the X-ray-selected AGN population, and likely have higher SFRs than the other half of the population.

On the other hand, \citet{Bundy08} studied the properties of the host galaxies of X-ray selected AGN 
at $0.4<z<1.4$ in the DEEP2/Palomar survey \citep{Bundy06}, which includes the EGS, and found a 
different result. They estimated the star formation quenching rate, defined as the 
number of galaxies that move to the red sequence per Gyr. They found that this quenching rate coincided with the AGN 
triggering rate, assuming an AGN lifetime of $\sim$1 Gyr. \citet{Bundy08} claimed that the agreement 
between the quenching and triggering rates may constitute an evidence of a physical link between the two
phenomena. However, the latter authors do not consider those X-ray selected AGN the cause of the  
quenching, but simply that they are somehow associated to it.

Thus, by putting together all the previous results, it seems that, for X-ray selected
AGN, the period of moderately luminous AGN activity may not have strong influence in the star 
formation activity of the galaxies. 
In this context, our result would be against previous evidence for
moderately luminous AGN not quenching star formation. However, the majority 
of these works are based on the assumption that
the bulk of FIR and submillimiter emission is dominated by the host galaxy (see \citealt{Mullaney11} and 
references therein), and they are then used as a proxy of the star formation activity. Although this may 
be a valid assumption in general, other authors have identified the narrow-line region clouds as 
the most likely location of the cool, FIR emitting dust
\citep{Dicken09}. In any case, such assumption has associated, to some degree, an overestimation the 
SFRs measured in AGN hosts. 

An alternative scenario would be the existence of a time delay 
between the offset of the star formation and nuclear activity, as observed 
in the local universe. \citet{Davies07} analysed the star formation in the nuclear region of nine
local Seyfert galaxies on scales of 10-100 pc and found evidence for recent, but not longer
active, star formation. Sampling larger scales (up to 2 kpc radius) of SDSS selected starburst galaxies, 
\citet{Wild10} found that the 
average rate of accretion of matter onto the black hole rises steeply $\sim$250 Myr after the onset 
of the starburst. A similar result was found by \citet{Schawinski09}. More recently, \citet{Hopkins12}
reported simulations of AGN fuelling by gravitational instabilities that naturally produce a delay 
between the peaks of SFR and nuclear activity. This offset scales as the gas consumption time and 
it is similar to those suggested by the observations on both small and large scales \citep{Davies07,
Schawinski09,Wild10}.

In order to confirm/discard the previous hypotheses, it is of extreme importance to perform 
accurate estimations of the AGN component 
to obtain reliable measurements of the SFRs in AGN hosts at $z\sim1$ and beyond. In the future, 
we aim to repeat this study for a larger sample of AGN to derive
statistically significant results. Considering the typical J-band magnitudes
of AGN at redshift $z\sim0.8$ ($\sim$20 mag), the use of 8/10 m telescopes is necessary to 
obtain higher signal-to-noise H$\alpha$ detections. The new/upcoming NIR instruments FLAMINGOS-2
on the 8 m Gemini-South and EMIR on the 10 m Gran Telescopio Canarias (GTC) will represent a 
definitive leap in the study of intermediate-to-high redshift AGN.

%On the other hand, more luminous AGN (L$_{2-10 keV} \ga 10^{44}~erg~s^{-1}$), 
%and higher SFRs would be linked via galaxy mergers/interactions \citep{Lutz10,Santini12}.

%We have estimated reliable SFRs uncontaminated from AGN emission using NIR spectroscopy for the first 
%time for a sample of active galaxies at this redshift, and compared them with those obtained from 
%their 24 \micron~emission. The median SFR obtained using the two methods is 6 M$_{\sun}~yr^{-1}$, which is 
%considerably lower than the $\sim$30 M$_{\sun}~yr^{-1}$ measured for non-active 
%star forming galaxies at $z\sim$1 and M$_* \ga 10^{10.5}M_{\sun}$ (\citealt{Noeske07,Twite11}; Rodr\' iguez-Eugenio 
%et al.~in prep.).

\section{Conclusions}

We present new NIR spectroscopic observations of a representative sample of 28 X-ray and MIR selected sources
in the EGS with a median redshift
of $z\sim0.8$ ($\Delta$z=[0.3, 1.3]). These galaxies show a wide variety of SED shapes, 
that we use to divide the sample in AGN-dominated and host galaxy-dominated. We combined LIRIS NIR spectra with 
DEEP2 optical spectroscopic data to maximize the number of 
H$\alpha$ and H$\beta$ detections. The main results from this study are summarised as follows:

\begin{itemize}

\item The NIR spectra (rest-frame optical) of the sample show a wide variety of spectral features, including 
prominent emission lines typical of AGN, H$\alpha$ and H$\beta$ in absorption, 
weak emission lines or featureless spectra, and emission lines with double kinematic components.

\item For 89\% of the sample, the spectroscopic and SED classifications are in agreement, 
confirming the reliability of multifrequency SED fits to classify X-ray and MIR sources at intermediate redshift. 

\item Using different diagnostics, we can confirm the presence of nuclear activity in 24/28 sources (86\%). 
The X-ray and MIR emission of the remaining four galaxies can be produced either by a heavily obscured low-luminosity AGN or 
intense star formation.

\item We estimate the AGN contribution to the observed 24 \micron~emission using the SED fits 
used to classify the galaxies: 97\% for {\it Type-1}, 72\% for 
{\it Type-2}, 88\% for {\it SB-cont}, 4\% for {\it SB-dom}, and 25\% for {\it NG}.

\item We calculate SFRs from attenuation and AGN-corrected H$\alpha$ luminosities and obtain values within the 
interval SFR = [0.03, 19] M$_{\sun}~yr^{-1}$, which are in good agreement 
with the SFRs obtained from observed 24 \micron~fluxes: SFR = [0.1, 166] M$_{\sun}~yr^{-1}$. The average ($\pm$standard deviation) 
and median SFRs independently obtained from the two methods are (7$\pm$7, 7) M$_{\sun}~yr^{-1}$ and (20$\pm$50, 5) 
M$_{\sun}~yr^{-1}$ respectively.

\item By comparing our results with those published in the literature for non-active star-forming galaxies of similar
stellar masses and redshifts, we find that our SFRs are lower on average, although with a large dispersion. 
Despite the small size of the samples involved 
in this comparison, as well as the uncertainty affecting our AGN-corrected SFRs, 
the results provide an indication that the presence of an AGN in a galaxy at $z\sim0.8$
might be quenching its star formation. Alternatively, we might be 
seeing a delay between the offset of the star formation and AGN activity, as observed in the local universe.

\end{itemize}

\appendix
\section{DEEP optical spectra}
\label{appendix}

\begin{figure*}
\centering
{\par
\includegraphics[width=12cm]{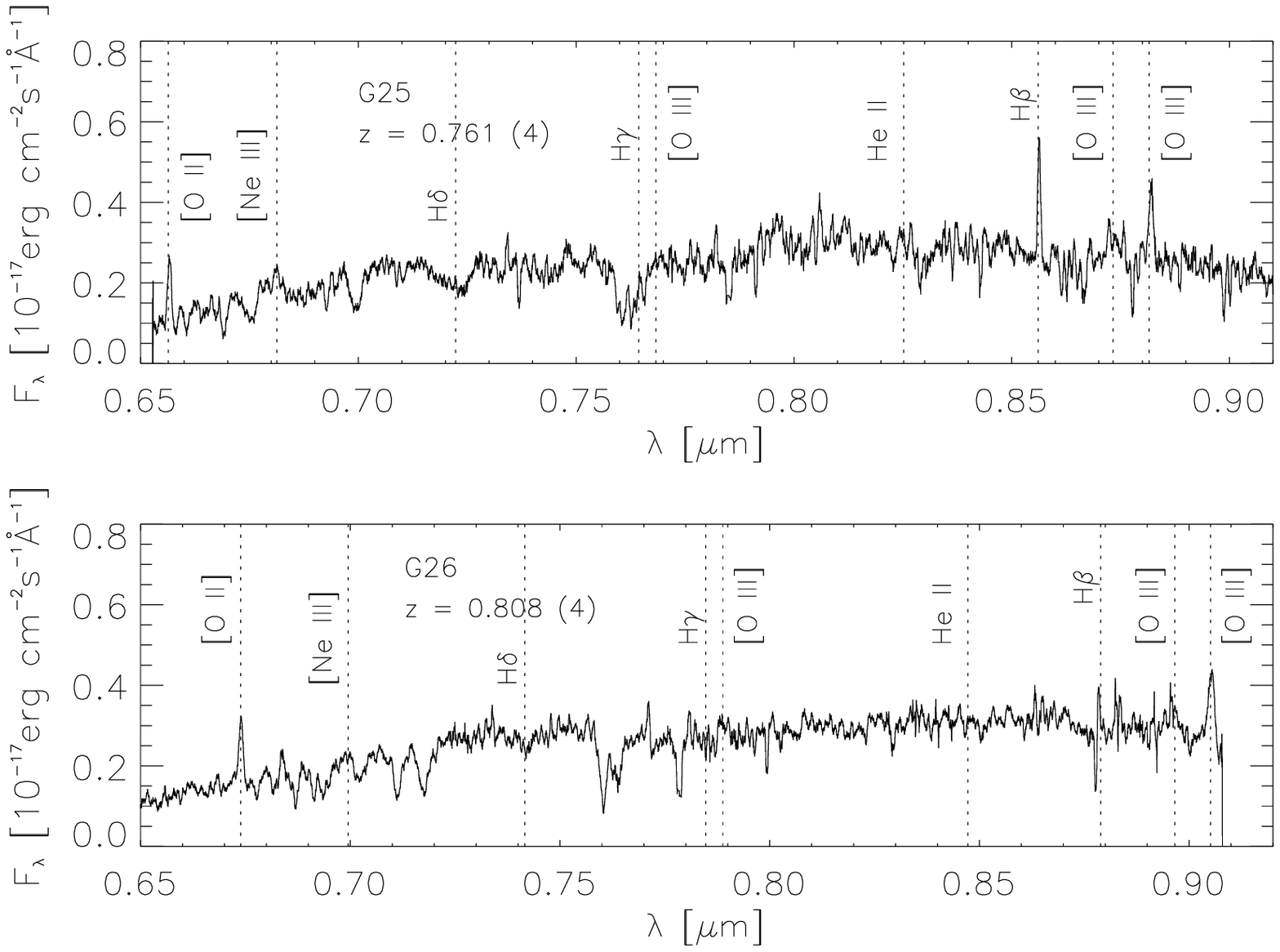}
\includegraphics[width=12cm]{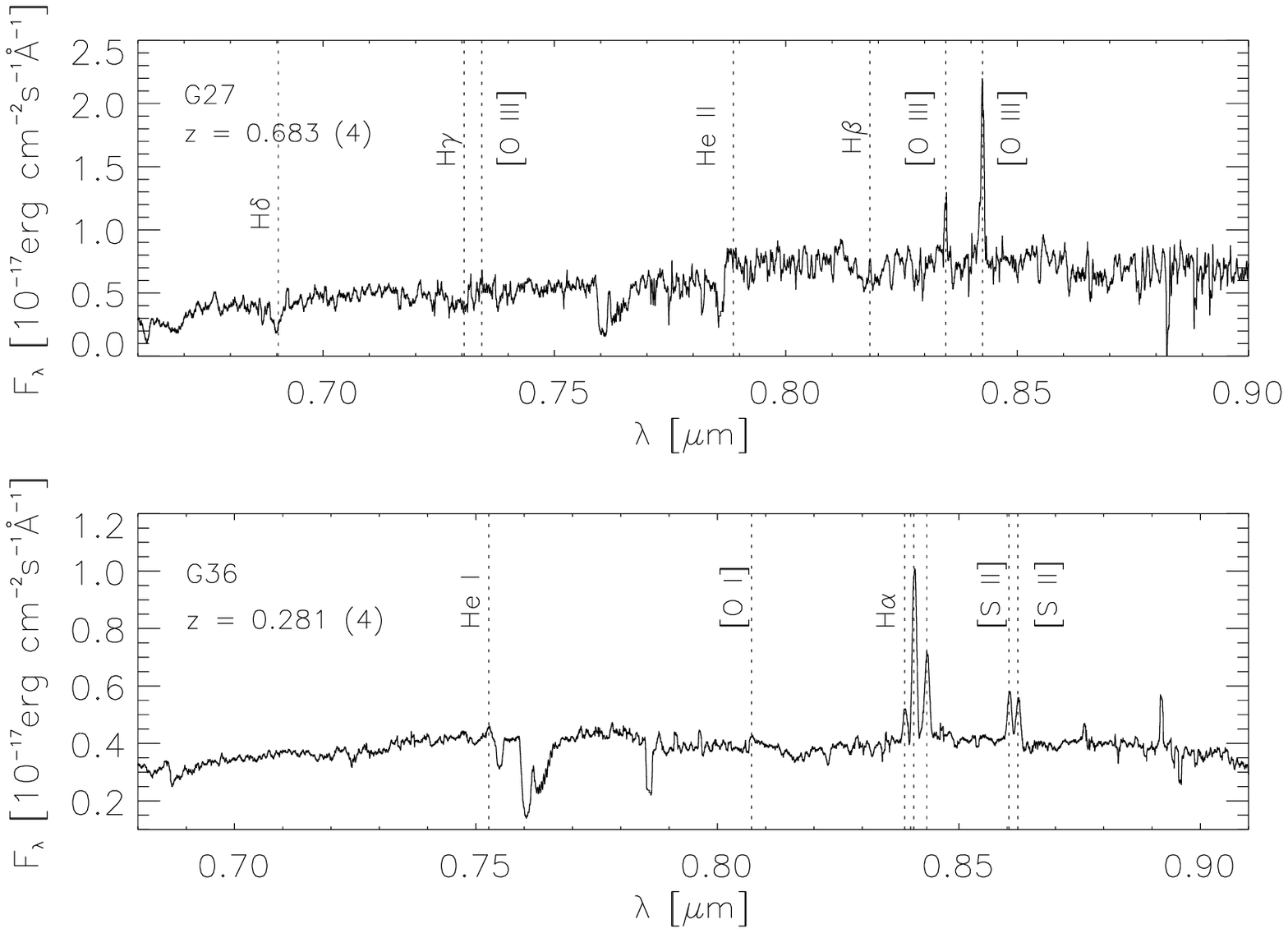}\par}
\caption{DEEP2 optical spectra of the galaxies G25, G26, G27, and G36. 
Typical AGN emission lines are labelled. The H$\alpha$ labels correspond to H$\alpha$+2[N II].
The DEEP2 spectroscopic redshift is labelled in each panel, together with its reliability between brackets (1-2 = low and 
3-4 = high reliability). {\bf The optical spectra of all the galaxies in the sample are available in the electronic version of the 
journal.}}
\label{deep1}
\end{figure*}

Figures \ref{deep1} to A5 show the optical spectra of the 20 galaxies in the sample with
available data from DEEP2 \citep{Davis03}. For details on how the observations were performed and other
technical details, we refer the reader to Section \ref{opt}. The flux calibration of these spectra
was done by scaling them to the individual NIR LIRIS spectra.

\section*{Acknowledgments}

The authors acknowledge the Spanish Ministry of Science and Innovation (MICINN) through project 
Consolider-Ingenio 2010 Program grant CSD2006-00070: First Science with the GTC 
(http://www.iac.es/consolider-ingenio-gtc/). 
C.R.A. acknowledges the Estallidos group through project PN AYA2010-21887-C04.04 and 
STFC PDRA (ST/G001758/1). 
A.A.H acknowledges support from the Spanish Plan Nacional de Astronom\' ia y Astrof\' isica under grant 
AYA2009-05705-E and from the Universidad de Cantabria through the Augusto Gonz\' alez Linares Program. 

The authors acknowledge Kevin Schawinski, Roser Pell\'o, Carlos Gonz\'alez Fern\'andez, Omaira Gonz\'alez Mart\' in, Guillermo Barro, 
Andrew Cardwell, Berto Gonz\'alez, and Juan Carlos Guerra for their valuable help.

The authors acknowledge the data analysis facilities provided by the Starlink Project, 
which is run by CCLRC on behalf of PPARC.

Based on observations made with the William Herschel Telescope operated on the island of La Palma by 
the Isaac Newton Group in the Spanish Observatorio del Roque de los Muchachos of the Instituto 
de Astrof\' isica de Canarias under the CAT programs 18-WHT7/07B, 16-WHT7/08A, 26-WHT11/08B, and 75-WHT23/09A. 

Funding for the DEEP2 survey has been provided by NSF grants AST95-09298, AST-0071048, AST-0071198, 
AST-0507428, and AST-0507483 as well as NASA LTSA grant NNG04GC89G.

Some of the data presented herein were obtained at the W. M. Keck Observatory, which is operated as 
a scientific partnership among the California Institute of Technology, the University of California 
and the National Aeronautics and Space Administration. The Observatory was made possible by the 
generous financial support of the W. M. Keck Foundation. The DEEP2 team and Keck Observatory 
acknowledge the very significant cultural role and reverence that the summit of Mauna Kea 
has always had within the indigenous Hawaiian community and appreciate the opportunity to 
conduct observations from this mountain. 

We finally acknowledge useful comments from the anonymous referee.

\label{lastpage}

\end{document}